\documentclass[11pt,oneside]{article}
\usepackage{setspace,graphicx,amssymb,amsmath,latexsym,amsfonts,amscd,amsthm,multirow,ctable,mathdots,caption,array,diagbox,mathtools}
\usepackage{chet}
\usepackage{tabularx,cite,mathrsfs}
\usepackage{authblk}
\usepackage{color}
\usepackage{tcolorbox}

\usepackage{fullpage}
\usepackage{stmaryrd}
\usepackage{rotating}

\newcommand{\eq}[2] {\begin{equation} \label{#1} #2 \end{equation}}
\def\eqr{\eqref}

\newcommand{\be}{\begin{equation}}
\newcommand{\ee}{\end{equation}}
\newcommand{\bea}{\begin{align}}
\newcommand{\eea}{\end{align}}
\newcommand{\nn}{\nonumber}
\newcommand{\bb}{\mathbb}
\newcommand{\mc}{\mathcal}
\def\({\left(}
\def\){\right)}
\newcommand{\half}{\frac{1}{2}}
\renewcommand{\a}{\alpha}
\renewcommand{\b}{\beta}
\renewcommand{\d}{\delta}

\newcommand{\m}{\mu}
\newcommand{\n}{\nu}

\renewcommand{\th}{\theta}
\newcommand{\D}{\Delta}

\newcommand{\Tr}{{\rm Tr \,}}

\newcommand{\es}[2] {\begin{equation} \label{#1} \begin{split} #2 \end{split} \end{equation}}

\newcommand{\mat}[2]{\left(\begin{array}{#1} #2 \end{array}\right)}
\newcommand{\tr}{\mathrm{Tr}}

\def\Z{\mathbb{Z}}

\def\C{\mathbb{C}}
\def\c{{c\over 24}}
\def\Tr{ {\rm Tr~}}
\def\Oc{{\cal O}}

\def\p{\partial}
\def\eps{\epsilon}

\def\D{\Delta}

\def\p{\partial}

\def\vac{{\rm vac}}

\newcommand{\cali}{\includegraphics[height=3.725mm, trim = 0mm 0mm 0mm 0mm, clip]{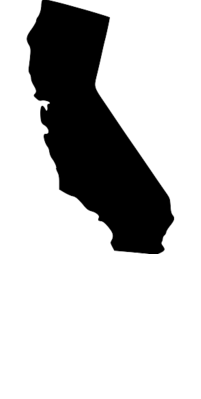}}
\newcommand{\mass}{\includegraphics[height=3.725mm, trim = 0mm 0mm 0mm 0mm, clip]{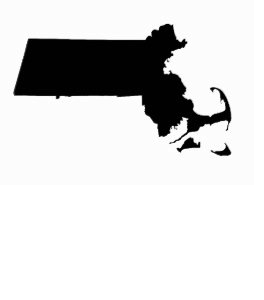}}
\newcommand{\nj}{\includegraphics[height=4.225mm, trim = 0mm 0mm 0mm 0mm, clip]{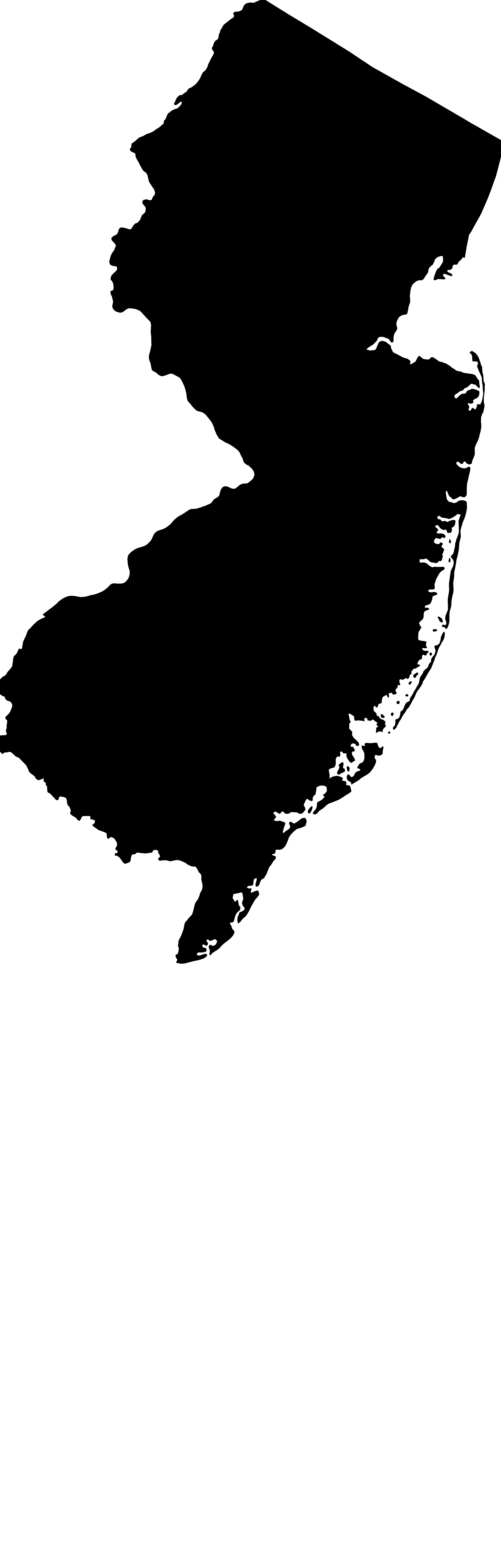}}
\newcommand{\canada}{\includegraphics[height=3.725mm, trim = 0mm 0mm 0mm 0mm, clip]{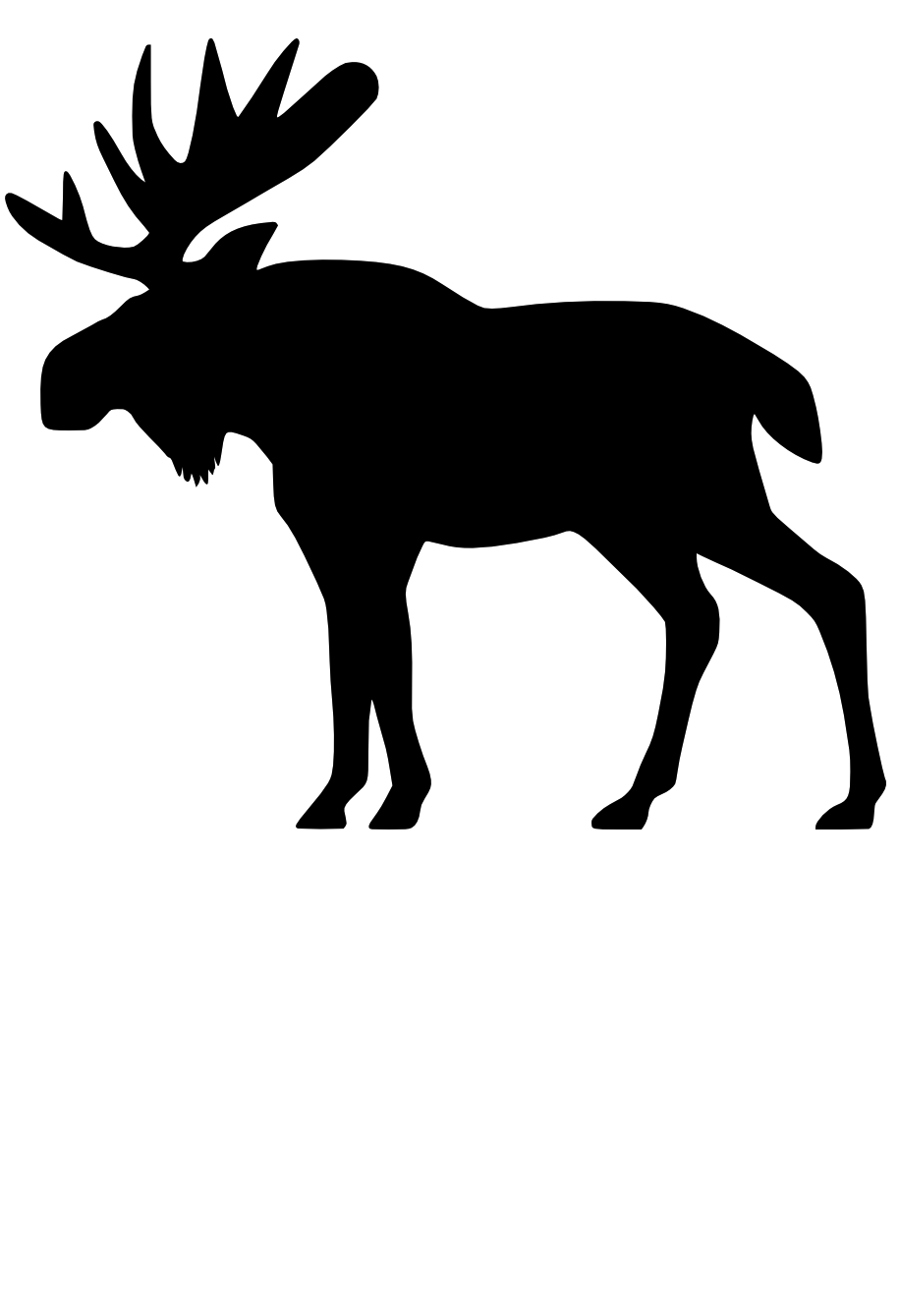}}

\usepackage{hyperref}

\numberwithin{equation}{section}

\begin{document}


\begin{center}

{
~ \\ ~ \\ ~\\ ~\\ ~\\ ~\\~
\Huge Small Black Holes and Near-Extremal CFTs}
\end{center}
\vspace{0.5cm}
\begin{center}
         {Nathan Benjamin\cali,}
           {Ethan Dyer\cali},
                  {A. Liam Fitzpatrick\mass},
                  {Alexander Maloney\canada},
          {Eric Perlmutter\nj}\\

\vspace{0.5cm}\cali {\it Stanford Institute for Theoretical Physics, Via Pueblo, Stanford, CA 94305, USA}\\

\mass {\it Boston University Physics Department, Commonwealth Avenue, Boston, MA 02215, USA}\\

\canada {\it McGill Physics Department, 3600 rue University, Montr\'eal, QC H3A 2T8, Canada}\\

\nj {\it Department of Physics, Princeton University, Princeton, NJ 08544, USA}\\

\end{center}

\begin{center}
  {\bf Abstract} 
\end{center}

Pure theories of AdS$_3$ quantum gravity are conjectured to be dual to CFTs with sparse spectra of light primary operators. The sparsest possible spectrum consistent with modular invariance includes only black hole states above the vacuum. Witten conjectured the existence of a family of extremal CFTs, which realize this spectrum for all admissible values of the central charge.  We consider the quantum corrections to the classical spectrum, and propose a specific modification of Witten's conjecture which takes into account the existence of ``small" black hole states. These have zero classical horizon area, with a calculable entropy attributed solely to loop effects. 
Our conjecture passes various consistency checks, especially when generalized to include theories with supersymmetry.  In theories with $\mc{N}=2$ supersymmetry, this ``near-extremal CFT'' proposal precisely evades the no-go results of Gaberdiel et al.
\newpage
\tableofcontents

\section{Introduction and Summary}
\label{sec:intro}

The AdS/CFT Correspondence \cite{Maldacena:1997re,GKP,Witten} states that every theory of gravity in Anti-de Sitter space (AdS) is dual to a conformal field theory (CFT) which lives on the boundary of AdS.  Not every CFT, however, is dual to a semi-classical theory of gravity.  This can be understood most precisely in AdS${}_3$/CFT${}_2$, where the CFT central charge is equal to the AdS radius in Planck units \cite{Brown:1986nw}
\be
c={3\ell_{\text{AdS}} \over 2G_N}.
\ee
We must therefore require that $c$ is large if we want to have a semi-classical bulk dual.

This constraint is not enough, however.  Light operators in the CFT correspond to perturbative bulk fields in AdS, so one must
require that the CFT should not have too many light operators if the bulk theory is to remain under perturbative control.  
It is interesting, therefore, to attempt to find CFTs whose spectrum of light states is as sparse as possible. 
Such CFTs can be regarded as UV complete theories of AdS gravity which have the property that the regime of validity of semi-classical pure Einstein gravity 
extends to as high energy as possible.  
Optimistically, one might hope to find a ``pure" theory of gravity which contains only graviton degrees of freedom below the Planck scale.  This was the strategy proposed by Witten \cite{Witten:2007kt}.\footnote{More generally, there has recently been significant progress in understanding the emergence of semi-classical gravitational behavior purely from constraints directly on the spectrum and OPE coefficients in CFTs; an incomplete list is \cite{Heemskerk:2009pn,Fitzpatrick:2010zm,Headrick:2010zt,ElShowk:2011ag,Fitzpatrick:2012cg,Hartman:2013mia, Perlmutter:2013paa, Hartman:2014oaa,  Asplund:2014coa, Alday:2014tsa,Camanho:2014apa,Fitzpatrick:2014vua, Belin:2014fna, Haehl:2014yla},\cite{Asplund:2015eha, Benjamin:2015hsa,Benjamin:2015vkc,Hartman:2015lfa,Hofman:2016awc,Li:2015itl,Maldacena:2015iua,Perlmutter:2016pkf,Anous:2016kss}.}  Further motivation for studying this case is the apparent universality of many features of semi-classical gravity, independent of their matter content.  So pure gravity gives insights into the universal features of the gravitational sector of an any semi-classical gravity theory.

In AdS$_3$, such a constraint is particularly appealing since it would imply that the low-energy degrees of freedom include only boundary gravitons, whose spectrum and interactions are completely determined by the Virasoro symmetry.  Thus, at low energies the theory is completely fixed by its central charge.  At high energies, black hole states enter the spectrum. Their dynamics are not known to be fixed by symmetry and are notoriously difficult to understand. 
Indeed, it is not entirely clear that one can make sense of an AdS theory whose action is purely gravitational, or exactly where the first black hole states should enter in such a theory.  One can take as guidance semiclassical and perturbative calculations, but a satisfactory non-perturbative definition is lacking. 
The main focus of this paper is an improved quantitative estimate for the spectrum of black hole states in a pure gravity theory. 

As an example of the power of this technique, let us first consider a chiral conformal field theory, i.e. a theory with only right-moving degrees of freedom.\footnote{Such a theory would be dual not to pure Einstein gravity but instead to chiral gravity \cite{Li:2008dq}; the precise relationship with extremal CFTs was described in \cite{Maloney:2009ck}.}  Modular invariant, chiral CFTs have central charge $c=24 k$ with $k$ an integer, and all states have integer conformal weight $h$.  The vacuum energy is normalized to have $h=-k$. Modular invariance and holomorphicity then imply that the ``light'' or ``polar'' ($-k \le h< 0$) spectrum in such a theory completely determines the ``heavy'' ($h\ge 1$) spectrum.  In \cite{Witten:2007kt}, Witten conjectured that pure gravity should be described by a CFT whose light spectrum includes only gravitons, i.e. Virasoro descendants of the vacuum:
\be
Z_{k}(\tau) = q^{-k} \prod_{n=2}^\infty \frac{1}{1-q^n}+ {\cal O}(q)~.
\label{eq:extremalPF}
\ee
A theory with partition function (\ref{eq:extremalPF}) is known as an extremal CFT. 
At $k=1$, there exists a (possibly unique) theory with this partition function, which is the Monster CFT of Frenkel, Lepowsky and Meurman \cite{FLM}.  At $k\ge 2$, despite many attempts, no extremal CFTs have been found. 
This failure may reflect the non-existence of $k\ge 2$ extremal CFTs.\footnote{We refer the reader to \cite{Gaberdiel:2007ve, Gaiotto:2008jt, Gaberdiel:2008pr} for a discussion of potential constraints on extremal CFTs, and \cite{Dixon:1988qd, Duncan, Cheng:2014owa, Benjamin:2015ria, Harrison:2016hbq} for construction of some extremal CFTs with supersymmetry at low central charge.}  It may, however, simply reflect the fact that the space of chiral CFTs becomes much more complicated as one increases $k$.  For example, while at $c=24$ there are (under reasonable assumptions) only 71 possible chiral CFTs \cite{Schellekens:1992db}, at $c=48$ the number of possible theories is at least $10^{120}$.

A key motivation for Witten's extremal CFT conjecture was that the lightest BTZ black hole has zero classical horizon area, and hence vanishing Bekenstein-Hawking entropy; by AdS/CFT, this maps to an absence of primary states at $h=0$. However, as already noted in \cite{Witten:2007kt}, the extremality proposal might be too strict.  In this paper, we will argue that condition (\ref{eq:extremalPF}) is too extreme because the massless BTZ black hole receives quantum corrections to its entropy.
Specifically, we will show that already in gravitational perturbation theory, one can obtain an estimate for the number of black hole states at $h=0$, and that it is generically positive and grows exponentially with $\sqrt{k}$.  Correspondingly, in constructing the partition function of the CFT dual to pure gravity, the extremal CFT partition function \eqref{eq:extremalPF} should be supplemented by an additional constant:
\be
Z_k(\tau) = q^{-k} \prod_{n=2}^\infty \frac{1}{1-q^n}+ {\cal O}(q^0)~.
\label{eq:nearextremalPF}
\ee
We call such theories ``near-extremal CFTs.'' More generally, one might use this term to denote any CFT whose spectrum is extremal, plus a small number of operators near threshold; such a CFT would still be dual to pure gravity in a a semi-classical limit. In this paper, we use the term narrowly, adding states only exactly at threshold. 

We will now describe our findings in more detail. We first list our gravity results according to the strength of our arguments, from least to most speculative; these are followed by an independent result derived in chiral CFT:

\subsubsection*{\it Quantum corrections to $S_{\rm BH}(h \ge 1)$}

For very massive black holes, which have $h\gg k$, the number of black hole microstates is given to good approximation by the Bekenstein-Hawking entropy $\frac{A}{4 G_N}$.  However, the entropies of ``small'' black holes, whose radius is small compared to the AdS curvature radius, have significant deviations from the classical formula.  We will show that these corrections are in fact comparable to the leading area law contribution for states near threshold.  In particular, for states with $1 \le h \ll k$ the loop corrections to the black hole entropy become important. This correction takes the form
\be
S_{\rm BH}(h) \approx 4 \pi \sqrt{k \left( h + \frac{1}{24} \right)},
\label{eq:introshift}
\ee
at large $k$, where the $\frac{1}{24}$ is absent in the classical formula for the entropy.  At $h \sim {\cal O}(1)$, this is not a small correction: it is parametrically the same size as the classical entropy, and clearly larger than logarithmic corrections that are usually considered. This is very similar to the corrections to black hole entropy for four dimensional string compactifications described in e.g. \cite{Sen:1995in, LopesCardoso:1998tkj, Dabholkar:2004yr, Dabholkar:2004dq, Sen:2004dp, Hubeny:2004ji, Dabholkar:2005dt}.

In fact, one can see that this correction must be present, by looking at states just above threshold. 
As already noted, imposing that the polar spectrum of the gravitational partition function is populated only by boundary gravitons fixes the black hole spectrum above $h=0$ due to the constraints of modularity and holomorphicity of the partition function. An unambiguous computation of the resulting entropy of primaries at low values of $h\ge 1$ is then possible, and shows that the above correction to the area law is necessary; see Figure \ref{fig:n0h1}. Similar results hold in ${\cal N}=1$ and ${\cal N}=2$ supergravity.

\subsubsection*{\it Quantum corrections to $S_{\rm BH}(0)$: Perturbative conjectures}

The limit $h \rightarrow 0$ corresponds to the limit where the classical black hole radius vanishes, and the entire entropy can be thought of as coming from quantum effects.  At $h=0$, the number of states is not uniquely fixed by the number of states at $h<0$ and the constraints of modular invariance. Moreover, it is not clear that one can trust perturbative calculations around the semi-classical background with zero size horizon. Nevertheless, we will find that the entropy $S_{\rm BH}(h)$ that is valid for $h\ge 1$ can be computed to all orders in $1/k$ and has a smooth continuation to $h=0$.  In the cases of pure non-supersymmetric gravity, 
the resulting entropy is
\be
S_{\rm BH}(0) = \log \left[2\sqrt{3} \left( e^{\frac{\pi}{6} \sqrt{24 k-1}} - e^{\frac{\pi}{6} \sqrt{24k-25}} \right)\right] .
\label{eq:introSN0}
\ee
We conjecture that this is the correct number of states up to non-perturbatively small corrections in $k$ (i.e. corrections of the form $e^{-\alpha k}$ for some $\alpha$). 

As a strong check of this conjecture, we perform the analogous computation in ${\cal N}=1$ supergravity, for which we obtain
\be
S_{\rm BH}(0) = \log \left[ \sqrt{2} \left( e^{\pi \sqrt{ \frac{k^*}{2} - \frac{1}{16}} } - e^{\pi \sqrt{\frac{k^*}{2} - \frac{9}{16}}}\right) \right] \qquad ({\cal N}=1)~,
\label{eq:introSN1}
\ee
where $k^*\equiv \frac{c}{12}$ is an integer.
In this case, there is a bound due to Witten \cite{Witten:2007kt} on the number of black hole states at $h=0$.  It is an upper bound for $k^*$ odd and a lower bound for $k^*$ even.  At large $k^*$, it varies smoothly as a function of $k^*$ and can therefore be thought of as rapidly oscillating between being an upper bound and a lower bound.  Assuming the actual entropy $S_{\rm BH}(0)$ is a smooth function of $k$ at large $k$, the bound completely fixes $S_{\rm BH}(0)$ up to non-perturbatively small corrections.  We find that this prediction exactly matches (\ref{eq:introSN1})!

At ${\cal N}=2$, a similar result holds, and one can perform additional checks as well.  In this case, the superconformal algebra contains an $R$-current $J$, so one can consider charged black holes with $R$-charge $\ell$.  The classical black hole radius is controlled by the quantity $h-\frac{3\ell^2}{2c}$. A powerful constraint is that the graded Ramond sector partition function with $(-1)^F$ inserted must be a weak Jacobi form.  The analysis of \cite{Gaberdiel:2008xb} used this fact to argue that at large $c$, one is {\it forced} to have additional ``black hole'' states for negative values of the classical black hole radius, in the regime $-\frac{1}{8} < h- \frac{3 \ell^2}{2c} < 0$.  We calculate the correction to the black hole entropy at $h> \frac{3 \ell^2}{2c}$ and find a shift similar to the $1/24$ in (\ref{eq:introshift}):
\be
S_{\rm BH}(h,\ell) = 2\pi \sqrt{ \left( \frac{c}{6} - \frac{1}{2} \right) \left( h - \frac{3 \ell^2}{2 c} + \frac{1}{8} \right)}.
\ee
Extrapolating into the regime $h< \frac{3 \ell^2}{2c}$, this predicts that there should be states exactly in the window where the analysis of \cite{Gaberdiel:2008xb} found that they are required.

\subsubsection*{\it Quantum corrections to $S_{\rm BH}(0)$: Non-perturbative conjectures}

Finally, one would ideally like to predict the non-perturbatively small corrections in $S_{\rm BH}(0)$ and determine its value exactly. Any such conjecture must of course give a non-negative integer, and furthermore in a large $k$ expansion it should agree with the ``perturbative'' conjectures (\ref{eq:introSN0}), (\ref{eq:introSN1}).\footnote{For ${\cal N}=1$, this agreement must hold since, as we show, (\ref{eq:introSN1}) can be uniquely fixed by the bound in \cite{Witten:2007kt}.  For the non-supersymmetric case, this constraint is only as strong as the evidence in support of (\ref{eq:introSN0}).}  We explore a number of possible conjectures, but ultimately find none of them to be fully satisfactory.  Nevertheless, we find tantalizing hints of structure that could point the way towards a complete conjecture for all values of $c$. 

Our efforts (in Section \ref{secva}) can be briefly summarized as follows. First, we analytically continue the gravitational Rademacher sum, which uniquely determines the density of states with $h>0$ given the perturbative spectrum, to $h=0$. This can be done in a very natural way that yields a non-negative integer for all $k$.\footnote{In the non-supersymmetric case at $k=1$, it yields $N_{\rm BH}(0)=24$ rather than zero, which indeed lies in the discrete set of allowed values at $k=1$ \cite{Schellekens:1992db}. This would uniquely select the Leech lattice CFT as the holographic dual, rather than the Monster CFT.} However, its large $k$ expansion appears not to agree with the more rigorous determination of the perturbative expansion leading to \eqr{eq:introshift}. Taking another tack, we then analyze the generating function of partition functions in the space of chiral CFTs whose polar spectra match that of pure gravity. This does not lead to a sharp conjecture for $N_{\rm BH}(0)$, but shows that $N_{\rm BH}(0)$ is directly related to the modularity properties imposed on such an object. Modular transformations in the space of theories relate small and large radius gravity, so this would be a fascinating route to a complete conjecture.

Optimistically, the partition function of pure gravity should be determined by a sum over geometries.  While in pure Einstein gravity the sum over geometries appears problematic \cite{Maloney:2007ud, Keller:2014xba}, the sum can be made precise in the case of chiral gravity \cite{Maloney:2009ck}.  The torus partition function is then computed in terms of a sum over 3-manifolds with torus boundary.  The sum over torus handlebodies gives an expression for the partition function in terms of a Rademacher sum.  One interesting possibility is that it might be necessary to include additional saddles in this sum which modify only the constant term in the partition function.  Such geometries were already identified in \cite{Maloney:2007ud}, where it was noted that saddles with cusps solve the local equations of motion.  Although we will not attempt to compute their contribution to the path integral precisely, since the loop corrections are difficult to compute, these saddles would naively contribute only to the constant term in the partition function.

\subsubsection*{\it Small black holes from modularity in chiral CFT}

In Section \ref{secv}, we briefly turn to the CFT side. We do not attempt to explicitly construct near-extremal CFTs, largely due to the absence of a sharp non-perturbative conjecture for the number of states at $h=0$ at all $k$. However, by combining the powers of holomorphy and modularity, we are able to prove the following: in a chiral CFT with $c=24k$ and at least one spin-1 current, the number of Virasoro primaries at level $k$ above the vacuum, call it $N_k$, is bounded from below, with a lower bound that grows exponentially with $\sqrt k$. At large $k$, and to leading order, 
\eq{nkboundintro}{N_k\gtrsim  e^{2\pi\sqrt{{c_{\rm eff}k\over 6}}}}
where $c_{\rm eff}$ counts the number of light primaries in the CFT. The derivation utilizes modular properties of the current two-point function on the torus. This may be viewed as a CFT derivation of small AdS black hole degeneracies, and supports the perspective that our pure gravity calculations do describe the gravitational sector of more general theories containing small amounts of matter. A tempting extrapolation of this result would be that {\it all} sparse theories at large $k$ must have an exponential density of threshold primaries, whether or not they contain spin-1 currents; this would, among other things, rule out large $k$ extremal CFTs. We leave a proper look into this speculation for future work. 
\vskip .2 in
The remainder of the paper is organized as follows. In Section \ref{sec:N0}, we describe the case of pure gravity.  In Sections \ref{sec:n1} and \ref{sec:n2}, we generalize to ``pure'' ${\cal N}=1$ and ${\cal N}=2$ supergravity, respectively. In Section \ref{secva}, we discuss the possibility of extending the perturbative expressions for the partition function of pure gravity theories into an exact, non-perturbative conjecture. Finally, in Section \ref{secv}, we derive a constraint on the number of threshold primaries from chiral CFT. Appendices collect various calculational details supplementing the text, and some modular background material.

\section{Non-supersymmetric Near-Extremal CFTs}
\label{sec:N0}

\subsection{Classical Action}

To calculate the entropy of states at a given energy, one may attempt to compute the torus partition function $Z(\tau)$ by evaluating the path integral for the action
\be
I = \frac{1}{16 \pi G_N} \int d^3 x \sqrt{ g} \left( R +  \frac{2}{\ell_{\rm AdS}^2} \right)
\ee
 with boundary conditions given by a fixed modular parameter $\tau$ (see e.g. \cite{Kraus:2006wn} for a review). 
Saddle points of the action are quotients of AdS$_3$ \cite{Carlip:1994gc,Maldacena:1998bw, Maloney:2007ud}.  Subleading contributions come from boundary gravitons, which can be neglected at leading order in large $k$ (with $h/k>0$ fixed).  The black hole solution is related to the vacuum by a modular transformation $\tau \rightarrow -1/\tau$, so in fact the leading entropy is given by taking the contribution $q^{-k}$ of the vacuum to the partition function, performing the modular transformation, and taking a Laplace transform.  Since this procedure is identical to the derivation of the Cardy formula, the results are identical as well:
\be
S_{\rm BH}(h) = 4 \pi\sqrt{k h} + 4 \pi \sqrt{k \bar h}= \frac{A_{\rm BH}(h)}{4 G_{\rm N}}.
\label{eq:BekHawk}
\ee
At $h=0$, the Schwarzschild radius, 
\be\label{eq:srad}
r_+ = \frac{\ell_{\rm AdS}}{2}  \sqrt{h\over k} = 8\sqrt{h k}\, G_N\,,
\ee
and therefore the classical entropy of the black hole, vanishes. This was part of the motivation to set the number of black hole states in pure gravity to zero at $h=0$ in \cite{Witten:2007kt}.
Of course, even perturbatively (\ref{eq:BekHawk}) has quantum corrections, and it should have non-perturbative corrections as well. Therefore, as noted in \cite{Witten:2007kt}, it may not be necessary to take the prediction (\ref{eq:BekHawk}) so literally.

\subsection{Quantum Corrections}

Before considering the one-loop correction to the black hole entropy, it is useful to review the one-loop correction to the partition function for the AdS vacuum, since the latter is relatively transparent and the two are related by a modular transformation.  The exact gravitational partition function for the (thermal) AdS vacuum saddle plus boundary excitations is essentially fixed by Virasoro symmetry to equal the Virasoro vacuum character \cite{Maloney:2007ud},\footnote{Henceforth we focus only on the chiral half of all entropies, partition functions, and so on. We will note as necessary when a result is specific to chiral gravity.}
\eq{chivac}{\chi_{\vac}(\tau) = q^{-k}\prod_{n=2}^{\infty}{1\over 1-q^n}\equiv q^{-k}\sum_{h=0}^{\infty} d_h q^h}
where $d_h$ is the number of Virasoro descendants $h$ above the vacuum at level $-k$,
\eq{}{d_h \equiv p(h)-p(h-1)~,}
and $p(h)$ is the number of integer partitions of $h$. Taking the log of $Z_{\rm TAdS}=\chi_{\vac}$, it is easy to see its order-by-order structure in a $1/k$ expansion:
\be
\log Z_{\rm TAdS} = -k \log q -  \sum_{n=2}^\infty \log (1-q^n) .
\label{eq:N0VacChar}
\ee
There are no terms proportional to inverse powers of $k$, so the contribution from the vacuum saddle is one-loop exact.  

The perturbative calculation of black hole entropies has an analogous form. Given any partition function of the form
\eq{}{Z(\tau)= \sum_{h'=-k}^{\infty} C_{h'}q^{h'}~,}
one performs a modular transformation, then a Laplace transform, to obtain the microcanonical BTZ black hole entropy \cite{Maloney:2007ud}. Whereas the semi-classical result follows from a modular transformation of the vacuum contribution, the loop contributions also receive contributions from the vacuum descendants.  The final form is 
\be
N_{\rm BH}(h) = e^{S_{\rm BH}(h)} = 2\pi \sum_{h'=-k}^\infty C_{h'} \sqrt{\frac{-h'}{h}} I_1(4 \pi \sqrt{-h h'}),
\label{eq:OneLoop}
\ee  
In pure gravity,
\be
C_{h'} = d_{h'+k}\,, \ \ \ \ \ h^{\prime}\,<\, 0
\label{eq:DescFromPart}
\ee
Taking just the vacuum $h'=-k$ contribution in (\ref{eq:OneLoop}) and expanding at large $h$ or $k$ obtains 
\be
S_{\rm BH}(h) \approx 4 \pi \sqrt{h k} + \frac{1}{4} \log \frac{k}{4} - \frac{3}{4} \log h + \dots.
\ee

The naive expectation is that contributions in (\ref{eq:OneLoop}) with $h'>-k$ are exponentially suppressed due to the Bessel function factor $I_1(4\pi\sqrt{-hh'}) \sim e^{4 \pi \sqrt{|h h'|}}$.  However, the coefficients $C_{h'}$ are also growing exponentially, and for small $h/k$ this growth overwhelms the Bessel function decay.   Using the Hardy-Ramanujan formula for $p(n)$, we can approximate $N_{\rm BH}(h)$ as
\be
N_{\rm BH}(h)  \sim \sum_{ h'=-k}^0 e^{2\pi \sqrt{\frac{1}{6}\(h'+k\) - \frac{1}{144}}} e^{4\pi\sqrt{h |h'|}}
\label{eq:nokloos}
\ee
where we approximated the number of Virasoro descendants at level $x$ by its Cardy growth, $e^{2\pi \sqrt{\frac{1}{6} x-\frac{1}{144}}}$. We then approximate (\ref{eq:nokloos}) by saddle point. In particular, the saddle occurs at
\be
h' = -\frac{h (24 k-1) }{1+24h}.
\label{eq:newsaddle}
\ee
In the limit of large $h$ and $k$ with $h\gtrsim k$, the saddle point therefore occurs at $h' \approx -k + \frac{k}{24 h} + \dots \approx -k$, so indeed the contribution from the vacuum dominates.  However, when $h/k$ is small, the saddle point moves away from $h' \approx -k$
and gives
\be
S_{\rm BH}(h)\approx 4\pi \sqrt{\left( k- \frac{1}{24} \right) \(h+\frac{1}{24}\)}, \qquad (h \ll k)
\label{eq:doitright}
\ee
so the argument of the square root is shifted.  At large $h,k$, this shift clearly produces a correction that is ${\cal O}(1/h, 1/k)$ suppressed compared to the leading term (\ref{eq:BekHawk}).  However, 
when $h \sim {\cal O}(1)$, the correction produced by this shift is comparable to the leading term even at large $k$, and so there is no reason to neglect it. This is one of our main results. 

Recall that the classical Schwarzschild radius in AdS$_3$ for an extremal spinning black hole is given by (\ref{eq:srad}), where $\ell_{\rm AdS}$ is the AdS curvature radius.  So the correction can be large even for black holes that are much larger than the Planck radius, $\ell_P \sim G_N$.   Black holes with $h < k$ are sometimes called ``enigmatic'' black holes and can have other unusual thermodynamic properties \cite{Denef:2007vg,Hartman:2014oaa,Gauntlett:2004wh}, which we comment on in Section \ref{sec:Thermo}.

\begin{figure}[t!]
   \centering
    \includegraphics[width=0.75\textwidth]{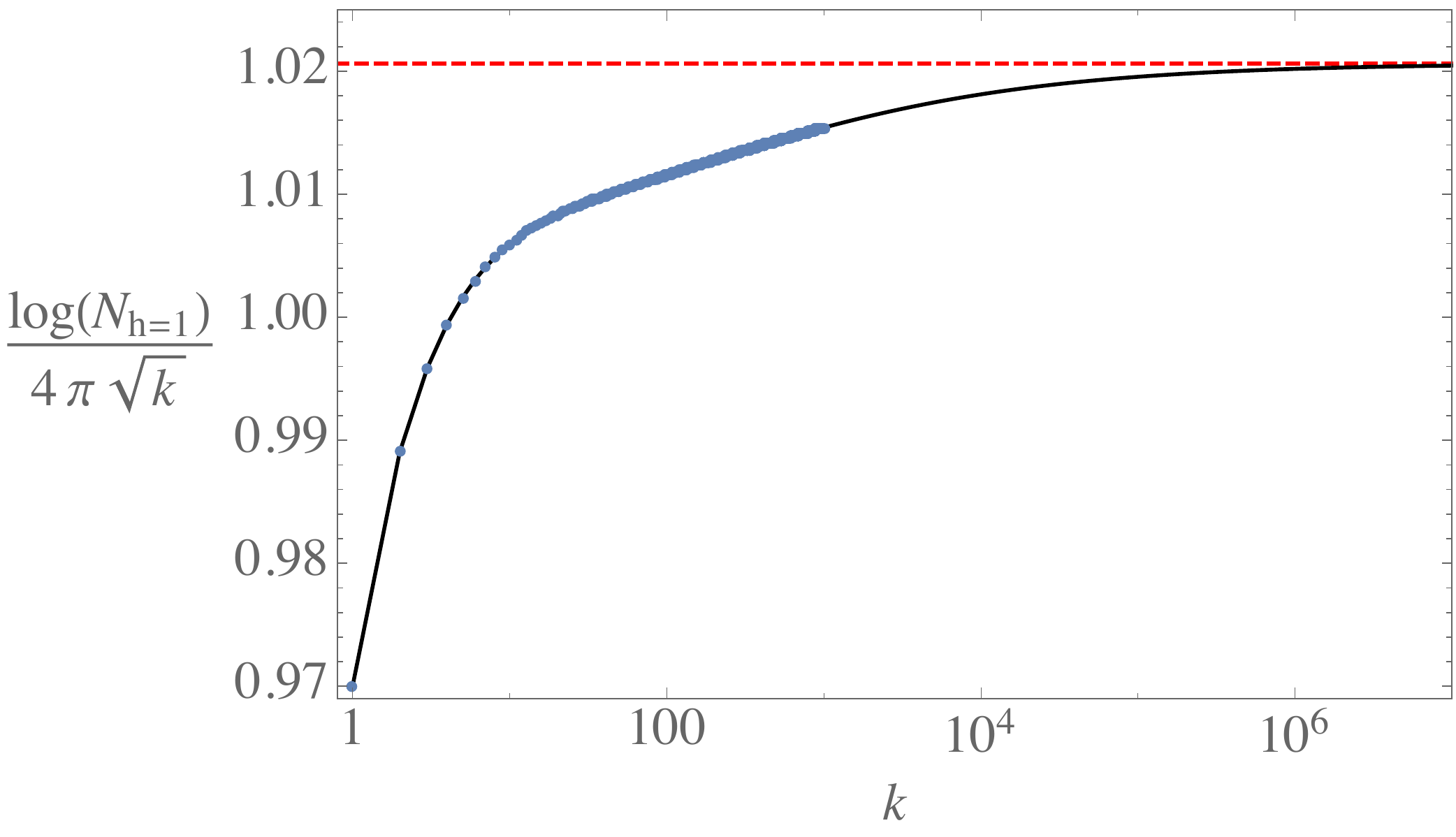}
    \caption{
    {\it Blue dots}: Ratio of the exact black hole entropy at $h=1$ to the prediction of Bekenstein-Hawking, computed up to $k=1000$.  For the first few values, $k=1,2,3,4$, $N_{h=1}$ appears to be approaching the Bekenstein-Hawking prediction, but continuing to higher $k$ one sees that $N_{h=1}$ actually overshoots it. 
    {\it Black solid}: The same ratio approximated with (\ref{eq:OneLoop}), with the sum cut off at $0$. This corresponds to taking only the perturbative part of the entropy at large $k$, and is an extremely good approximation.
    {\it Red dashed}: The asymptotic value of the ratio at large $k$, $\sqrt{\frac{25}{24}} \approx 1.02$; see (\ref{eq:doitright}).}
 \label{fig:n0h1}
\end{figure}

 In fact, for $h > 0$,  we can unambiguously see that this correction is required.  The reason for this is that, as mentioned in Section \ref{sec:intro}, once we have fixed the spectrum for $h < 0$, the spectrum at $h>0$ is {\it uniquely} determined by modular invariance in chiral theories. Thus the exact expression for $N_{\rm BH}(1)$, for example, can be determined. The result is that as $k$ becomes large, the approximation (\ref{eq:doitright}) approaches the correct value for $S_{\rm BH}(1)$, whereas the classical result (\ref{eq:BekHawk}) diverges from it (see Figure \ref{fig:n0h1}).

It seems natural, then, to take (\ref{eq:doitright}) as a one-loop-corrected approximation for the entropy which is valid at large $k$ with $h$ fixed.  Crucially, because of the shift, the entropy (\ref{eq:doitright}) no longer vanishes at $h=0$, but rather predicts a positive number of states, growing exponentially with $\sqrt k$.  We take this as suggestive evidence that the definition of CFTs dual to ``pure gravity'' should be modified to allow black holes at $h=0$, with entropy roughly given by (\ref{eq:doitright}). We call these ``near-extremal CFTs.'' 
 
 Higher order corrections to $S_{\rm BH}(0)$ can be included in an expansion in $1/k$ along the lines of Appendix \ref{app:N1Sub}, where we obtain the following expression which resums these perturbative corrections to all orders: 
 \begin{eqnarray}
 S_{\rm BH}(0) &=& \log \left[2\sqrt{3} \left( e^{\frac{\pi}{6} \sqrt{24 k-1}} - e^{\frac{\pi}{6} \sqrt{24k-25}} \right)\right] \nn\\
  &=&  2 \pi \sqrt{\frac{k}{6}} - \frac{1}{2} \log k + \log \(\sqrt{2}\pi\) + {\cal O}(k^{-1/2}) .
  \label{eq:N0log}
 \end{eqnarray}
 
 It is worth briefly noting the structure of subleading corrections when $h\gtrsim k$. There are two regimes. When $h/k\gg1$, we are in the Cardy regime of large black holes, and the leading correction from non-vacuum states comes from the first non-trivial descendant, with $h'=-k+2$. Its contribution is additive, and is suppressed compared to that of the vacuum by a factor of $e^{-4 \pi \sqrt{\frac{h}{k}}}\ll 1$. When $h/k\sim \Oc(1)$, the corrections from all descendants near the vacuum -- that is, with $h'\approx -k$ -- are parametrically comparable; while they do not shift the saddle, they do give constant, rather than exponentially suppressed, contributions to the entropy.

 We should comment on the upper bound in the sum on $h'$ in (\ref{eq:OneLoop}).  Since we have done the sum by saddle point, the result was independent of this upper bound.  However, it is clear that the expression for the sum formally stops making sense at $h' > 0$, since the integrand becomes imaginary.  One interpretation is that $h' \gtrsim 0 $ contributions are  exponentially small due to the Bessel function, and as such they are inseparable from other non-perturbative corrections.  This interpretation treats the sum over powers of $q$  somewhat analogously to the usual situation one faces with asymptotic expansions in powers of coupling constants. Indeed, this is what happens when computing the entropy of black holes above threshold \cite{Maloney:2007ud}. The contributions to $S_{\rm BH}(h>0)$ from terms in (\ref{eq:OneLoop}) with $h^{\prime}>0$ vanish. A non-perturbative formulation of the partition function in 3d gravity would presumably make this notion more precise for the states at $h=0$.

Note that if we define the quantum-corrected black hole threshold to be the value of $h$ at which $S_{\rm BH}(h)=0$, then the shift of $h$ pushes this down to $h=-1/24$. This is precisely the threshold between ``censored'' and ``uncensored'' states introduced in \cite{Keller:2014xba}. Our result can be viewed as a further evidence that the censorship threshold, not the polar threshold, determines the natural boundary between graviton and black hole states, up to log corrections.

 It is interesting to compare our result with earlier work on small black holes in string theory. In \cite{Sen:1995in, Dabholkar:2004yr, Dabholkar:2004dq}
 black holes which classically have zero horizon area were studied in type IIA string theory, where it was shown that quantum string effects generate a string-scale horizon, and an entropy proportional to its area. Those solutions are asymptotically flat, supersymmetric $d=4$ black holes, where the correction arises due to loop effects, the first of an infinite series of such corrections. Contrast this with our result, which requires no string theory or supersymmetry and gives the all-orders corrections to the entropy of asymptotically AdS$_3$ black holes due to graviton loops. Despite their differences, both settings do involve macroscopic corrections to solutions with classically zero entropy; in that sense, our computations are an AdS$_3$ analog of the higher-dimensional story.

\subsection{Phase Structure}
\label{sec:Thermo}

In the previous sections, we found that the entropy of black holes in pure gravity is approximately
\be
S_{\rm BH}(h) \approx 4 \pi \sqrt{ \hat{k} \left( h + \frac{1}{24} \right)} -\frac{1}{2}  \log \hat{k}, 
\ee
whereas the number of gravitons at level $h$ is 
\be
S_{\rm vac}(h) \approx 4 \pi \sqrt{  \frac{1}{24}\left( h + \hat{k} \right) } - \frac{3}{2} \log \hat{k},
\ee
where we have written the entropies in terms of a shifted central charge $\hat{k} \equiv k- \frac{1}{24}$. 
At $h=0$, these entropies are equal up to log corrections: $S_{\rm BH}(0) \approx S_{\rm vac}(0)$.\footnote{The entropy of black hole states is still logarithmically enhanced at $h=0$: $S_{\rm BH}(0) - S_{\rm vac}(0) \sim \log k$.} At large $k$ where $\widehat k \approx k$, in the absence of the shift $h \rightarrow h+1/24$, the number of gravitons would dominate up until $h= \frac{1}{24}$ instead. 
 \begin{figure}[t!]
   \centering
    \includegraphics[width=0.85\textwidth]{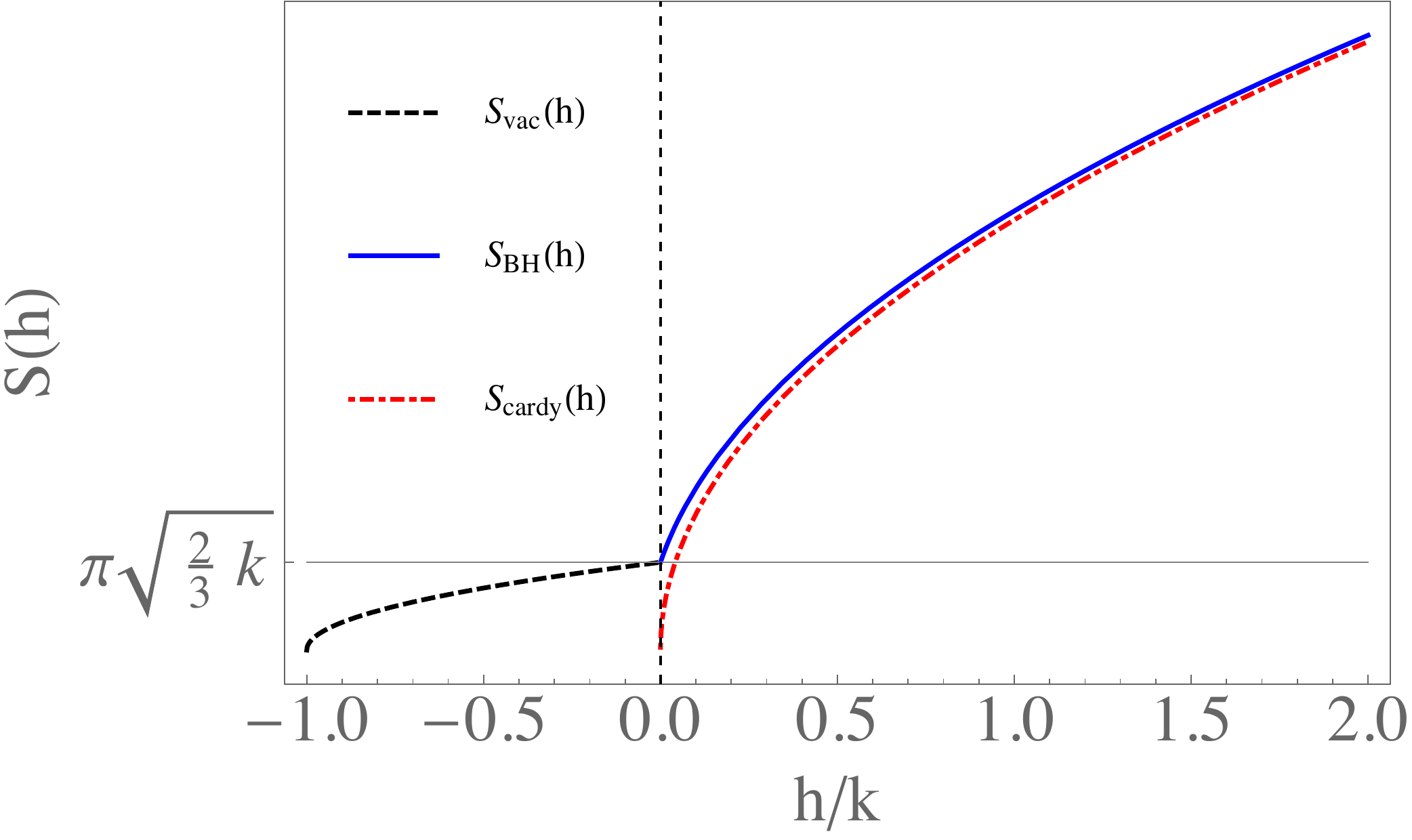}
    \caption{
    {\it Black, Dashed Line}: The leading order entropy due purely to vacuum gravitons.
    {\it Blue, Solid Line}: The entropy of black hole states including our shift. These meet at leading order in large $k$, $S_{\rm vac}(0)=S_{\rm BH}(0)=\pi\sqrt{2k/3}$.
    {\it Red, Dot-dashed Line}: The Cardy entropy, which vanishes at threshold.}
 \label{fig:entmiss}
\end{figure}
Because the leading pieces of $S_{\rm BH}(h)$ and $S_{\rm vac}(h)$ match at the threshold $h=0$ where the small black holes begin, in the microcanonical ensemble the effect of the shifts in $S_{\rm BH}$ is to smooth out the entropy as a function of $h$ somewhat: the discontinuity is only of ${\cal O}(\log k)$, not ${\cal O}(\sqrt{k})$. See Figure \ref{fig:entmiss}.  

In the canonical ensemble, however, the effect of the $h=0$ black holes is almost unnoticeable, since their contribution is subleading to that of the vacuum at all temperatures.  At large $k$, the Hawking-Page phase transition between the thermal vacuum and black holes occurs at the self dual inverse temperature, $\beta = 2 \pi$, and it is easy to see that the free energy for black holes at this point, $F(\beta=2\pi) \approx 4 \pi \sqrt{h k} -2 \pi (h+k)$, becomes positive only for $h\ge k$, which corresponds to $h-h_{\rm vac} = 2k$.  Thus, black holes with $h \lesssim k$ are thermodynamically unstable even though, due to the absence of Hawking radiation, there is no apparent dynamical instability.\footnote{In \cite{Hartman:2014oaa}, black holes in this regime were dubbed ``enigmatic'' black holes. This term originated in a more specialized context \cite{Denef:2007vg,Bena:2011zw}; we will use the looser definition which refers to any black hole configurations that never dominate the canonical ensemble at any temperature.} As discussed below \eqr{eq:N0log}, for black holes with $h\sim k$ the leading order saddle point does not shift; since these are the relevant states near the Hawking-Page transition, our results do not qualitatively change the phase structure of the theory at large $k$.

\section{$\mc{N}=1$ Near-Extremal CFTs}
\label{sec:n1}
In this section we generalize our counting of black hole states to supersymmetric theories. This allows us to put strict constraints on the number of states at threshold. We find that the conjecture for the number of $h=0$ black holes appropriately generalized to the $\mathcal{N}=1$ context obeys an infinite number of constraints.
\subsection{Classical Action and One-Loop Correction}

  It is conventional to parameterize the central charge in this case as $c=12 k^*$, where $k^*$ is an integer. The light spectrum now contains a fermionic excitation, the gravitino $\psi$, which appears in the supergravity action in AdS$_3$ as\footnote{See for example \cite{Deser:1982sw, Deser:1982sv, Hyakutake:2015qua}.}
\begin{eqnarray}
S &=& \frac{1}{16 \pi G_N} \int d^3 x \sqrt{g} \left[ R + \frac{2}{\ell_{\rm AdS}^2} - \bar{\psi}_\rho \gamma^{\mu \nu \rho} {\cal D}_\mu \psi_\nu \right. \nn\\
&& \left. 
+ \frac{\beta}{2} \epsilon^{\mu\nu \rho} \left(\omega_{\mu \ \b}^{\ a} \partial_\nu \omega_{\rho \ a}^{\ b} +\frac{2}{3} \omega_{\mu \ b}^{\ a} \omega_{\nu \ c}^{\ b} \omega_{\rho \ a}^{\ c}\right) - \frac{\beta}{2} \overline{D_\rho \psi_\sigma} \gamma^{\mu\nu} \gamma^{\rho \sigma} D_\mu \psi_\nu \right]. 
\label{eq:N1SugraAction}
\end{eqnarray}
The non-chiral case is $\beta=0$, and the (maximally) chiral case which we are considering is $\beta = \ell_{\rm AdS}$.
The gravitino modes are still purely boundary excitations 
and all created by fermionic generators $G_{-r}$ of the super-Virasoro algebra. The vacuum character of the $\mc{N}=1$ super-Virasoro algebra is
\be\label{supervac}
\chi_{\vac}^{{\cal N}=1}(\tau)= q^{-k^*/2}\prod_{n=2}^{\infty} \frac{1+q^{n-\half}}{1-q^n} \equiv  q^{-k^*/2}\sum_{\substack{n=0\\n\in\half\bb{Z}}}^{\infty} d_n^{\mc{N}=1}q^n ~.
\ee
The partition function for pure gravity should now contain these excitations in the light spectrum:
\be
Z(\tau)  =  \chi_{\vac}^{{\cal N}=1}(\tau) + {\cal O}(q^0).
\ee
Our main focus will again be the constant term, but as before we first consider the leading contribution to the entropy for general $h$.  The leading saddle point of the supergravity action is still obtained from thermal AdS by the modular transformation $\tau \rightarrow -1/\tau$, but now we have to include gravitino as well as graviton excitations,
\be
N_{\rm BH}^{\mc{N}=1}(h) = e^{S_{\rm BH}^{\mc{N}=1}(h)} = \pi \sum_{h'=-\frac{k^*}{2} \atop h' \in \frac{1}{2} \mathbb{Z} }^\infty C^{{\cal N}=1}_{h'} \sqrt{\frac{-h'}{h}} I_1(4 \pi \sqrt{-h h'}),
\label{eq:SugraOneLoop}
\ee  
 where now the coefficients $C_{h'}^{{\cal N}=1}$ for an extremal theory are the coefficients in the ${\cal N}=1$ vacuum character above:
 \be
 C_{h'}^{\mc{N}=1}=d_{h'+\frac{k^*}2}^{\mc{N}=1}\,, \ \ \ \ \ h^{\prime}<0\,.
 \ee
The derivation of the new $S_{\rm BH}(h)$ is just a slight variation of that in Section \ref{sec:N0}, but to clarify the character of the result more generally we will parameterize the growth of the coefficients $C_{h'}$ as
\be\label{vacgrow}
C_{-\frac{c}{24}+x} \stackrel{x \gg 1}{\sim} e^{2 \pi \sqrt{\frac{c_{\rm eff}}{6} (x-\frac{c_{\rm eff}}{24})}},
\ee
where $c_{\rm eff}=1,\frac{3}{2} $ for ${\cal N}=0,1$, respectively.  Eq. (\ref{eq:nokloos}) then generalizes to\footnote{The allowed values of $h'$ depend on the moding of the vacuum algebra generators, which we leave understood.}
\be
N_{\rm BH}(h) \sim \sum_{h'=-{c\over 24}}^{\infty} e^{2 \pi \sqrt{\frac{c_{\rm eff}}{6}(h'+\frac{c-c_{\rm eff}}{24})}} e^{4 \pi \sqrt{h |h'|}} ,
\label{eq:nokloosn1}
\ee
and the saddle point (\ref{eq:doitright}) shifts to yield
\be
S_{\rm BH}(h) \approx 4 \pi \sqrt{\frac{c-c_{\rm eff}}{24} \left( h + \frac{c_{\rm eff}}{24} \right)}, \qquad (h \ll \frac{c}{24}) .
\label{eq:generalceff}
\ee
We emphasize the generality of this result: it applies to any theory whose polar spectrum furnishes the vacuum representation of a chiral algebra with asymptotic growth of the form \eqr{vacgrow}.

Specializing again to ${\cal N}=1$, we have
\be
S_{\rm BH}^{\mc{N}=1}(h) \approx 4 \pi \sqrt{\left( \frac{k^*}{2}-\frac{1}{16} \right) \left( h + \frac{1}{16} \right) }, \qquad (h \ll k^*) 
\label{eq:n1bhentropy} 
\ee
At $h=0$, this gives approximately
\be
S_{\rm BH}^{\mc{N}=1}(0) \approx  \pi \sqrt{\frac{k^*}{2}}.
 \label{eq:blah}
\ee

These are our estimates for the leading parts of the quantum correction to the entropy of small black holes in ${\cal N}=1$ supergravity.  Additional corrections that are subleading in powers of $k$ can be  included with a more refined computation.  In Appendix \ref{app:N1Sub}, we compute these corrections and find
\begin{eqnarray}
S_{\rm BH}^{\mc{N}=1}(0) &=& \log \left[ \sqrt{2} \left( e^{\pi \sqrt{ \frac{k^*}{2} - \frac{1}{16}} } - e^{\pi \sqrt{\frac{k^*}{2} - \frac{9}{16}}} \right) \right]
\label{eq:preciselog} \\
&\approx& \pi \sqrt{\frac{k^*}{2}} - \frac{1}{2} \log k^*+ \log\left( \frac{\pi}{2} \right)+\ldots~.
\label{eq:blahlog}
\end{eqnarray}

\subsection{Comparison to a Bound of Witten}

\begin{figure}[t!]
\begin{center}
\includegraphics[width=\textwidth]{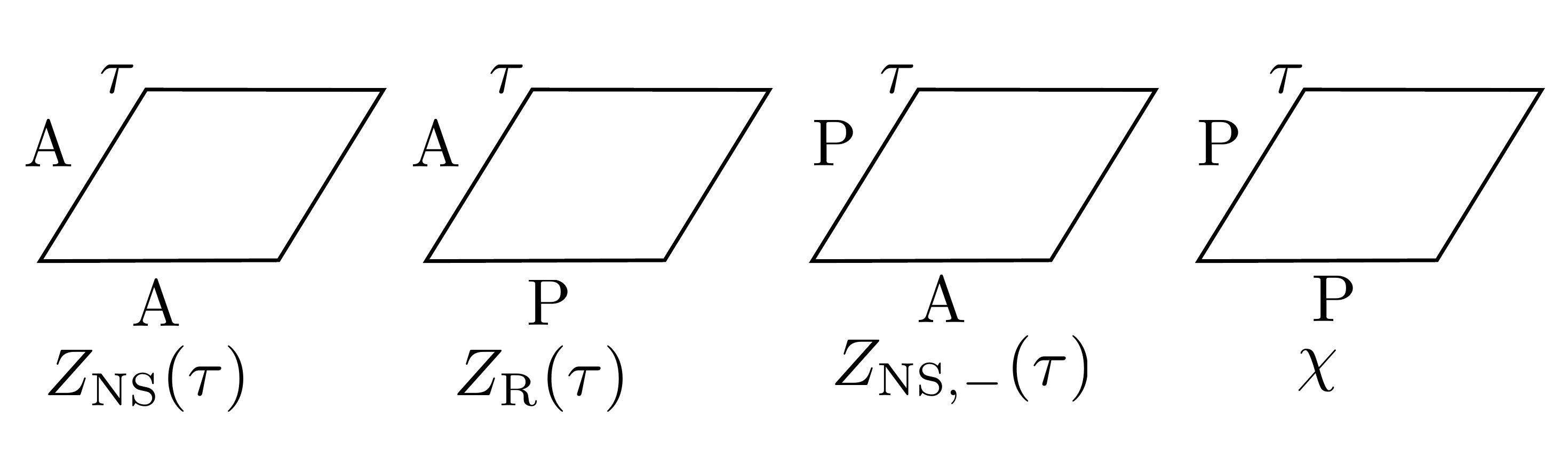}
\end{center}
\caption{The four possible spin structures on the torus.  ``A'' denotes that fermions have anti-periodic boundary conditions and ``P'' denotes periodic boundary conditions; the $x$-axis is the spatial direction. As noted, the first two spin structures correspond to the partition functions restricted to the Neveu-Schwarz or Ramond sectors, respectively.  The last two spin structures contain an additional $(-1)^F$ inside the trace; for the Ramond sector, this produces the Witten index $\chi = {\rm Tr}_{{\cal H}_{\rm R}} \left((-1)^F q^{L_0}\right)$.}
\label{fig:spinstructures}
\end{figure}

We next utilize a bound due to Witten \cite{Witten:2007kt} on the constant term in ${\cal N}=1$ theories to provide a powerful confirmation of \eqr{eq:preciselog}.  To begin with, we review the structure of states in an ${\cal N}=1$ CFT and the different spin structures on the torus.  
As we now have fermionic states, there are four possible spin structures on the torus, pictured in Fig. \ref{fig:spinstructures}.  Two of these correspond to the partition function in the Ramond (R) sector -- in which fermions are anti-periodic in the temporal direction but periodic in the spatial direction -- and the Neveu-Schwarz (NS) sector, in which fermions are anti-periodic in both directions: 
\begin{eqnarray}
&Z_{\text{NS}}(\tau) \equiv \tr_{\mc{H}_{\text{NS}}} q^{L_0}, \nn\\
&Z_{\text{R}}(\tau) \equiv \tr_{\mc{H}_{\text{R}}} q^{L_{0}}  ,
\label{eq:ns-}
\end{eqnarray}
where $L_0=-\frac{k^*}{2}=-\frac{c}{24}$ for the NS vacuum. 
Under modular transformations, these will generate a third spin structure under which fermions are periodic in the temporal direction but anti-periodic in the spatial direction, which corresponds to tracing over the NS sector with an additional factor of $(-1)^F$, i.e. 1 for bosons and $-1$ for fermions:
\be
Z_{\text{NS},-}(\tau)=\tr_{\mc{H}_{\text{NS}}} \left((-1)^{F}q^{L_{0}}\right).
\ee
The fourth and final spin structure $\chi$, 
\be
\chi=\tr_{\mc{H}_{\text{R}}} \left((-1)^{F}q^{L_{0}}\right),
\ee
with fermions periodic in both directions,  is just a constant counting the difference in the number of bosonic and fermionic Ramond sector ground states (i.e. the Witten index).  Given $Z_{\text{NS}},Z_{\text{NS},-}$, or $Z_{\text{R}}$, we can obtain the other two through 
\es{trtrans}{
Z_{\text{NS}}(\tau+1)&=(-1)^{k^{*}}Z_{\text{NS},-}(\tau),\\
Z_{\text{NS},-}(-1/\tau)&=Z_{\text{R}}(\tau)\, .
}
Each of these traces is invariant under a particular genus-zero subgroup of $SL(2,\mathbb{Z})$. $Z_{\text{NS}}$ is invariant under $\Gamma_{\theta}$; $Z_{\text{NS},-}$ is invariant under $\Gamma^{0}(2)$; and $Z_{\text{R}}$ is invariant under $\Gamma_{0}(2)$, where we define
\es{gt}{
\Gamma_{\theta}&=\left\{\mat{cc}{a&b\\c&d} \, \in \, SL(2,\mathbb{Z}) \ \ \ : a+b \textrm{ is odd and } c+d \textrm{ is odd}\right\}\\
\Gamma^{0}(2)&=\left\{\mat{cc}{a&b\\c&d} \, \in \, SL(2,\mathbb{Z}) \ \ \ : b \textrm{ is even}\right\}\\
\Gamma_{0}(2)&=\left\{\mat{cc}{a&b\\c&d} \, \in \, SL(2,\mathbb{Z}) \ \ \ : c \textrm{ is even}\right\}\,. 
}

Now we are ready to review the bound.  We demand that all the coefficients in both $Z_{\rm R}(\tau)$ and $Z_{\rm NS}(\tau)$ be non-negative. The transformations (\ref{trtrans}) imply
\be
Z_{\text{R}}(\tau) = (-1)^{k^*} Z_{\text{NS}}\(-\frac{1}{\tau}+1\).
\label{eq:modularrns}
\ee
This implies that the number of states in the Ramond sector at $L_0 = 0$ (corresponding to $q^0$ in $Z_{\text{R}}(\tau)$), call it ${\rm R}_0$, is given by (\ref{eq:modularrns}) in the limit $\tau \rightarrow i\infty$:
\be
\text{R}_0 \equiv Z_{\rm R}(i \infty) = (-1)^{k^*} Z_{\rm NS}(1).
\label{eq:R0vsPF}
\ee
Modular invariance does not prevent us from increasing or decreasing the number ${\rm NS}_0$ of NS sector primaries at $h=0$ by adding a constant $s$ to $Z_{\rm NS}$ without changing any of the other coefficients.  Note that by (\ref{eq:R0vsPF}), adding such a constant also shifts the number R$_0$ of Ramond ground states by $(-1)^{k^*}s$, but  leaves invariant the combination (first considered in  \cite{Witten:2007kt})
\be
\beta_{k^*} \equiv {\rm R}_0 - (-1)^{k^*} {\rm NS}_0 .
\label{eq:betadef}
\ee
Therefore, although the constant term in $Z_{\rm NS}$ is not fixed by its polar parts, the quantity $\beta_{k^*}$ {\it is} fixed by them.  In Appendix \ref{app:betaksGen}, we derive a generating function for $\beta_{k^*}$: 
\be
\sum_{k^* = 1}^\infty (-1)^{k^*}\beta_{k^*} q^{\frac{k^*}{2}} = \sqrt{\frac{\theta_3(\tau)}{\eta(\tau)^3}}\(\theta_4(\tau)^4-\theta_2(\tau)^4\)q^{\frac1{16}}\(1-\sqrt{q}\)
\label{eq:generatingbeta}
\ee
where we have used the standard Jacobi theta functions and Dedekind eta function (defined in Appendix \ref{app:thetabohnanza}).

Now, the key point is that both ${\rm R}_0$ and ${\rm NS}_0$ are non-negative, so the fact that $\beta_{k^*}$ is fixed by the polar parts of  $F(\tau)$ gives a bound on the number ${\rm NS}_0$ of NS sector primaries; for odd (even) $k^*$, this is an upper (lower) bound on ${\rm NS}_0$.
\es{betabound}{
\begin{array}{l l}
{\rm NS}_0>-\beta_{k^{*}}  & k^{*} \ \mathrm{even}\\
{\rm NS}_0<\beta_{k^{*}} & k^{*} \ \mathrm{odd}\,.
\end{array}
}
In principle, this gives us only a bound on ${\rm NS}_0$, but does not give us an estimate of its actual value.  In practice, though, the following feature of $\beta_{k^*}$ makes the bound significantly more informative:

\be
\textrm{For } k^* \geq 7,  \qquad \left\{ \begin{array}{cc} \beta_{k^*} >0 & k^* \textrm{ odd } \\ \beta_{k^*} < 0 & k^* \textrm{ even } \end{array} \right\} .
\ee
(If $k^* < 7$, $\beta_{k^*}$ is always positive.)

  Since $(-1)^{k^*+1} \beta_{k^*}$ is positive at large $k^*$ and oscillates rapidly between being an upper bound and a lower bound, it effectively becomes an approximate prediction for ${\rm NS}_0$. More precisely, asymptotically at large $k^*$ it has a series expansion of the form $\log |\beta_{k^*}| \sim a \sqrt{k^*} + b \log k^* + c + \frac{d}{\sqrt{k^*}} + \dots$. The saddle point calculation from gravity also has a series expansion of this form, and for these to agree an infinite number of series coefficients must match. So this is a highly non-trivial check for our result (\ref{eq:preciselog}).

\begin{figure}[t!]
   \centering
    \includegraphics[width=0.75\textwidth]{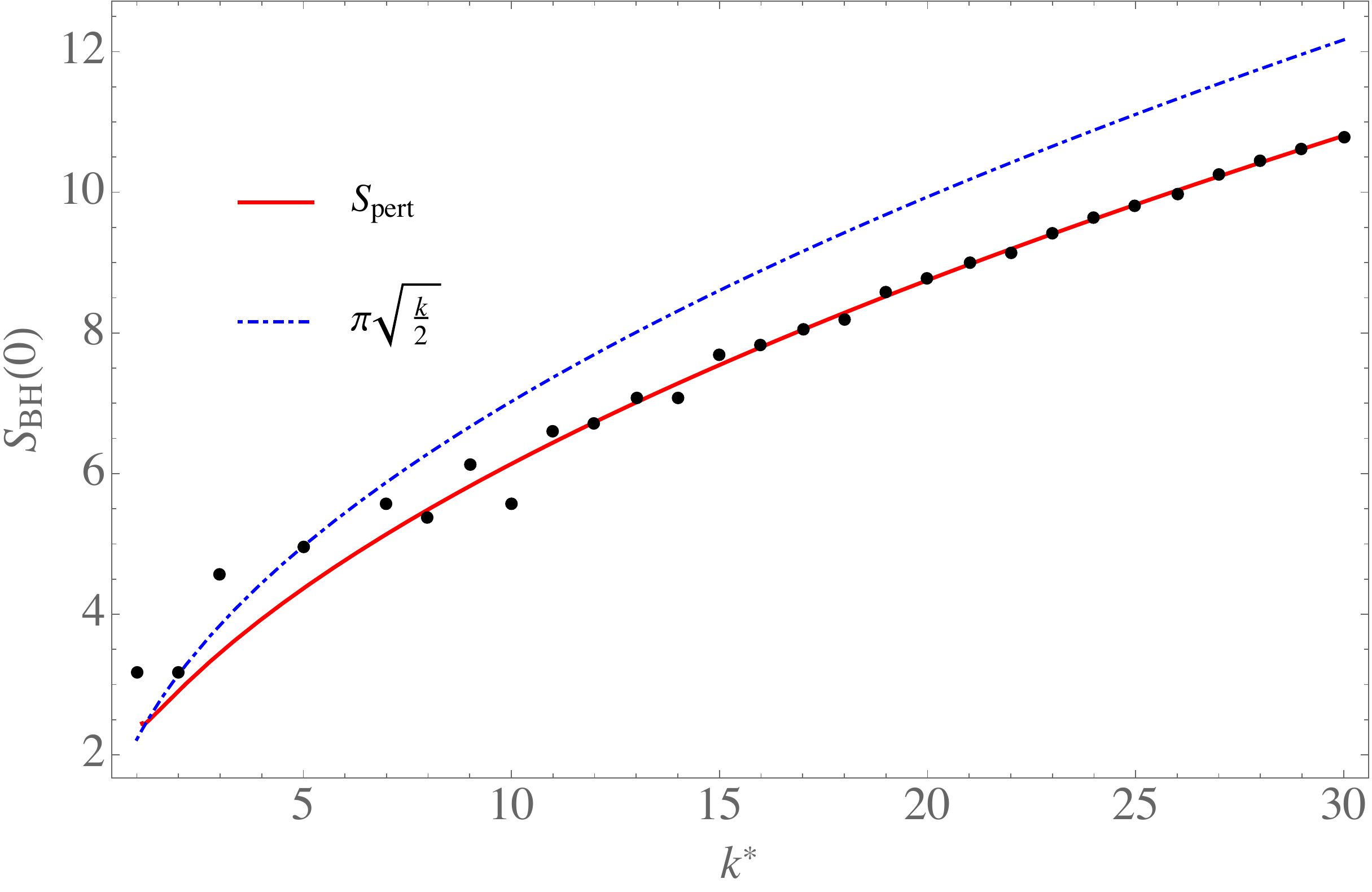}
    \caption{$h=0$ states for an $\mc{N}=1$ near-extremal CFT. {\it Red, solid}: Equation (\ref{eq:preciselog}), which is the all-orders in large $k^*$ expression from the saddle point calculation for $S_{\rm BH}(0)$.  
     {\it Black dots}: $\log{|\beta_{k^*}|}$. Note the strong agreement even for $k^* \sim 15$ - on this plot, the last few points at $k^* \sim 30$ appear to lie exactly on the asymptotic curve.  
    To show the trend in the size of the difference more clearly, we plot the relative error in Figure \ref{fig:n1} at large $k^*$.
     {\it Blue, dot-dashed}: $\pi \sqrt{\frac{k^*}{2}} $, the leading perturbative piece (\ref{eq:blah}), is also shown for comparison. } \label{fig:n1both}
\end{figure}

\begin{figure}[t!]
   \centering
    \includegraphics[width=0.75\textwidth]{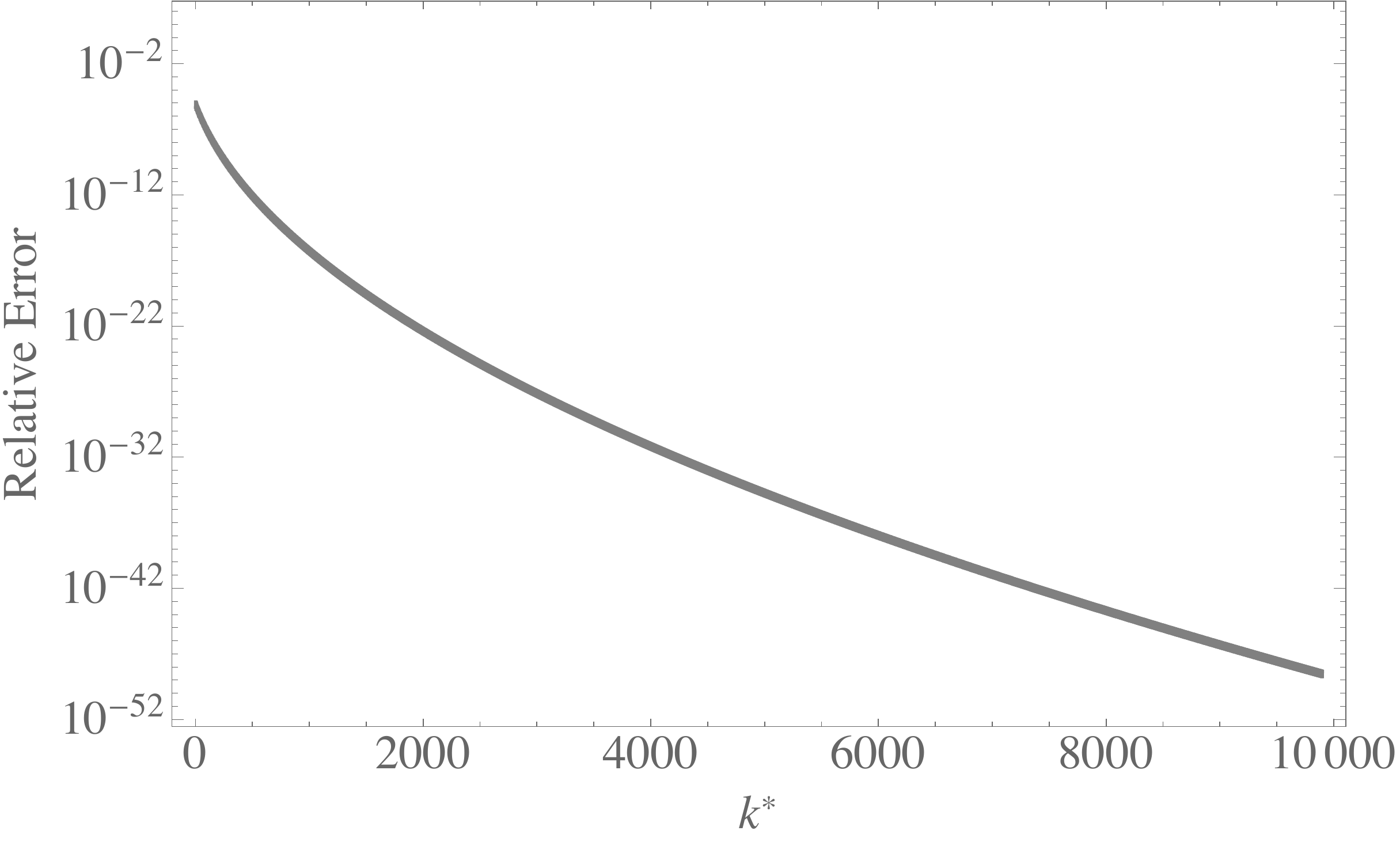}
    \caption{Here we plot the error 
    $1-\frac{\log{|\beta_{k^*}|}}{S_{\rm BH}(0)}$, with $S_{\rm BH}(0)$ from (\ref{eq:preciselog}), 
    as a function of $k^*$ for an $\mc{N}=1$ extremal CFT. The relative error decreases exponentially $\sim e^{-\frac{\pi}{2} \sqrt{\frac{k^*}{2}}}$, in accordance with the fact that all orders in the perturbative expansion are correctly reproduced by the gravity saddle point calculation. }
 \label{fig:n1}
\end{figure}

In Appendix \ref{app:MultSys}, we derive the asymptotics of $\beta_{k^*}$ at large $k^*$.  Remarkably, we find exact agreement with the result of the saddle point calculation (\ref{eq:preciselog})!
At large $k^*$, (\ref{eq:preciselog}) should give an exponentially accurate approximation for $\beta_{k^*}$, and indeed we see this agreement in Figures \ref{fig:n1both} and \ref{fig:n1}. 

We thus have an argument based solely on positivity in the Ramond sector partition function (and some smoothness conditions on it) that independently fixes the large $k^*$ behavior of the constant term in ${\cal N}=1$ supergravity to be (\ref{eq:blahlog}).

\subsection{Phase Structure}

Remarkably, the bound (\ref{betabound}) from unitarity can be understood in a completely different way, by demanding that theory has the correct phase structure for a bulk gravity dual.   To understand this,  note that if we interpret a partition function $Z(\tau)$, $Z_{\text{NS}}(\tau)$, etc. as the finite temperature partition function of the CFT on a circle, then  $\tau$ is interpreted as a complexified temperature
\be
\tau = \frac{1}{2\pi} \left(i \beta + \theta\right)
\ee 
with $\beta$ the usual (real) inverse temperature and $\theta$ a chemical potential conjugate to angular momentum.
All of the partition functions considered in this paper depend analytically on $\tau$ (or they are the absolute valued squared of something which depends analytically on $\tau$).
However, in the semi-classical limit, these partition functions should become non-analytic in $\tau$: they should exhibit the usual Hawing-Page phase transition between a thermal gas in AdS and the BTZ black hole.
This is a necessary condition for the bulk dual to describe semi-classical AdS gravity.
So the $Z(\tau)$ are a series of complex-analytic functions of $\tau$ which converge to a non-analytic function in the large central charge limit.  The important observation is that, since the $Z(\tau)$ are complex-analytic (rather than just real-analytic) functions of $\tau$, the way that this can happen is highly constrained.  In particular, the transition must occur through a condensation of zeros in the complex $\tau$ plane.  
In statistical mechanics this condensation of zeros is familiar in in the Lee-Yang description of phase transitions in lattice models.
The proposal that the Hawking-Page transition occurs via a condensation of Lee-Yang zeros was first described in  \cite{Maloney:2007ud}.  

In the case of AdS${}_3$ gravity, the phase transition occurs when the parameter $\tau$ exits the fundamental domain on the upper half-plane.  In particular, this means that we must require that the zeros of the partition function all lie along the unit circle in the complex $\tau$ plane at the edge of the fundamental domain so that in the large $k^*$ limit they condense into a branch cut corresponding to the Hawking-Page phase transition.  The locations of these zeros depend on the constant term, and it turns out that if ${\rm NS}_0$ is greater than (less than) $|\beta_{k^*}|$ for $k^*$ even (odd), at least one of these zeros does not lie on the unit circle.

To see why this is the case, begin by rewriting $Z_{\rm NS}$ in terms of the $\Gamma_\theta$-invariant function $K(\tau)$ \cite{Witten:2007kt}:\footnote{Our convention for $K(\tau)$ differs from that in \cite{Witten:2007kt} by the constant term of 24.}
\begin{align}
K(\tau) &\equiv \frac{\Delta^2(\tau)}{\Delta(2\tau)\Delta(\tau/2)} \nn \\
&= \frac{1}{\sqrt{q}} + 24 + 276\sqrt{q} + 2048q + \ldots, 
\end{align}
with $\Delta(\tau) = \eta^{24}(\tau)$ being the modular discriminant. The unit half-circle in $\tau$ (in the upper half-plane) maps to the interval $0 \le K(\tau) \le 64$. It is also convenient that 
\be
K(\tau=1)=0.
\ee  

The only poles of the partition function $Z_{\rm NS}(\tau)$ are at $\tau=i\infty\leftrightarrow K=\infty$, so $Z_{\rm NS}(\tau)$ is a polynomial in $K$.  Furthermore, $Z_{\rm NS}(\tau) = q^{-\frac{k^*}{2}} + \dots$, so it is a degree $k^*$ polynomial where the coefficient of $K^{k^*}$ is one.  Finally, its constant term is given by evaluating $Z_{\rm NS}(\tau)$ at $K=0$, in other words at $\tau=1$.  From equation (\ref{eq:R0vsPF}) and the definition of $\beta_{k^*}$ in (\ref{eq:betadef}), we therefore have
\be
 Z_{\rm NS}(K=0)=  \textrm{NS}_0+ (-1)^{k^*} \beta_{k^*}.
\ee
For $Z_{\rm NS}$ to have the desired phase structure, all of its roots $\{ K_i\}_{1 \le i \le k^*} = \{ K | Z_{\rm NS}(K)=0\}$  must lie in the interval $0 \le K \le 64$. In particular,  this implies the constant term of $Z_{\rm NS} = \prod_{i=1}^{k^*} (K-K_i)$ must be $(-1)^{k^*}$ times a positive number:
\be
0 \le (-1)^{k^*} Z_{\rm NS}(K=0) =  \textrm{NS}_0+ (-1)^{k^*} \beta_{k^*},
\ee
which reproduces the bound (\ref{betabound}).  

To state this conclusion another way, we see that the requirement that the number of Ramond ground states, $R_0$, is positive -- which is simply a statement of unitarity -- is also a consequence of the statement that the zeros of the partition function  lie on the phase boundary of the dual gravity theory.

\subsubsection*{Extension to non-supersymmetric case?}
Note that we can similarly demand that all the zeroes of the partition function lie on a boundary of a fundamental domain for the non-supersymmetric case. This would imply that the roots lie at $\tau = e^{i\phi}$ for $\frac{\pi}3 \leq \phi \leq \frac{\pi}2$, which corresponds to $0 \leq j(\tau) \leq 1728$ where $j(\tau) = \frac{1}{q} + 744 + \mc{O}(q)$. For all the roots to be real and positive in $j$, it is a necessary condition that when writing $Z(\tau)$ as a polynomial in $j$, the constant term be $(-1)^k$ times a positive number. Since $j(\tau)$ vanishes at $\tau=e^{\frac{i\pi}3}$, the constant term in the polynomial expansion of $j$ is given by $Z(e^{\frac{i \pi}3})$.

If we take for $Z(\tau)$ to be one of Witten's extremal CFT partition functions, it is not hard to see that
\be
Z(e^{\frac{i \pi}3}) \approx 3 \chi_{\text{vac}}(e^{\frac{i\pi}3})
\label{eq:approxbyvac}
\ee
where $\chi_{\vac}$ was defined in \eqr{chivac}. We will sometimes find it convenient to use the parameterization 
\be
\chi_{\text{vac}}(\tau) = \frac{q^{-k+\frac{1}{24}}}{\eta(\tau)}\(1-q\)
\label{eq:chivac}
\ee
Equation (\ref{eq:approxbyvac}) can be understood as follows.  Deep in the fundamental domain, near $\tau=i\infty$, we can approximate $Z(\tau)\approx \chi_{\text{vac}} (\tau)$.  Conversely, in any modular image of the fundamental domain under the image of some $\gamma = \left({a~b\atop c~d}\right) \in SL(2,\bb{Z})$, near the point $\tau = a/c$, we can approximate
$Z(\tau)\approx \chi_{\text{vac}} (\gamma\tau)$.
The point $\tau=e^{\frac{i\pi}3}$ is a fixed point under $TS\in SL(2,\bb{Z})$, which cubes to the identity.  It is the place where three modular images of the fundamental domain meet.  We can therefore approximate $Z(e^\frac{i\pi}3)$ by summing the vacuum character and its images under $TS$ and $(TS)^2$.  These all contribute equally, giving us a factor of $3$ in (\ref{eq:approxbyvac}). In the large $k$ limit, this approximation becomes infinitely accurate.\footnote{Similarly, one can approximate $Z(i)\approx 2 \chi_{\text{vac}}(i)$, since the vacuum character and its image under $S$ contribute equally giving a factor of $2$.}

Using (\ref{eq:chivac}), we see that
\be
Z(e^{\frac{i \pi}{3}}) \approx 3\alpha (-1)^k e^{\sqrt{3}\pi k}
\label{eq:tofuhouse}
\ee
where
\be
\alpha = (1+e^{-\sqrt{3}\pi}) \frac{e^{-\frac{\pi}{24}\(-i+\sqrt{3}\)}}{\eta(e^{\frac{i\pi}3})} \approx 1.00002.
\label{eq:alphago}
\ee
We can add a constant term to (\ref{eq:tofuhouse}) so long as the sign remains $(-1)^k$. For even $k$, we see that adding a constant term does not change the sign and so we get no constraint. For odd $k$, we get an upper bound for how many states we can add at $q^0$ (e.g. $744$ for $k=1$), that grows as $3\alpha e^{\sqrt{3}\pi k}$. Our prediction (\ref{eq:N0log}) satisfies (but does not saturate) this upper bound. Indeed, this growth is much faster than the ${\cal O}(e^{\sqrt{k}})$ growth characteristic of the sparse spectrum of 3d gravity, and so adds no constraint at all.\footnote{Said another way, no near-extremal CFT can saturate this bound: the total number of states at ${\cal O}(q^0)$ must be less than the number at ${\cal O}(q)$, and the latter grows as $e^{4\pi\sqrt{\frac{25}{24}k}}\ll e^{\sqrt{3}\pi k}$.} 

Note that the reason the non-supersymmetric case gave us a weak (exponential) bound whereas the $\mc{N}=1$ case gave a strong (subexponential) bound is because the fundamental domain of $\Gamma_\th$ reaches all the way down to $\text{Im~}\tau=0$, whereas for $SL(2,\bb{Z})$, the lowest value of $\text{Im~}\tau$ is $\frac{\sqrt{3}}2$.

\section{$\mc{N}=2$ Near-Extremal CFTs}
\label{sec:n2}

Moving up to $\mc{N}=2$ superconformal symmetry allows even further consistency checks of our conjecture.  In particular, it is now possible to grade the partition function with the $U(1)_\text{R}$ symmetry of the algebra.

We first study in subsection \ref{sec:ungrade} the ungraded partition function, using a similar analysis as in the previous section. In subsection \ref{sec:nogo}, we move on to the graded partition function, and review a no-go theorem found in \cite{Gaberdiel:2008xb}. Finally in subsection \ref{sec:winggkmo}, we show how our proposal precisely evades the no-go theorem.

\subsection{Ungraded $\mc{N}=2$}
\label{sec:ungrade}

The near-extremal partition function in the NS sector,
\be
Z_{\text{NS}}(\tau) = \Tr_{\mc{H}_{\text{NS}}}q^{L_0}~,
\ee
is defined so that its polar part matches that of the $\mc{N}=2$ vacuum character.  This character, which counts (ungraded) the  $\mc{N}=2$ Virasoro descendants of the vacuum, is
\be
\sum_{\substack{n=0\\n\in\half\bb{Z}}}^{\infty} d_n^{\mc{N}=2} q^n = \frac{1-\sqrt{q}}{1+\sqrt{q}} \prod_{n=1}^{\infty} \frac{(1+q^{n-\half})^2}{(1-q^n)^2}
\ee
so that
\be
Z_{\text{NS}}(\tau) = q^{-\frac{k^*}2} \frac{1-\sqrt{q}}{1+\sqrt{q}} \prod_{n=1}^{\infty} \frac{(1+q^{n-\half})^2}{(1-q^n)^2} + \mc{O}(q^0)
\label{eq:n2desc}
\ee
where $k^*=\frac{c}{12}$, as in Section \ref{sec:n1}. Since the modularity properties of $Z_{\rm NS}$ are the same as in the $\mc{N}=1$ case, we get a very similar expression to (\ref{eq:SugraOneLoop}) for the black hole entropy
\be
N_{\rm BH}^{\mc N=2}(h) = e^{S_{\rm BH}^{\mc N=2}(h)} = \pi \sum_{h'=-\frac{k^*}{2} \atop h' \in \frac{1}{2} \mathbb{Z} }^\infty C^{{\cal N}=2}_{h'} \sqrt{\frac{-h'}{h}} I_1(4 \pi \sqrt{-h h'}),
\ee
where the only difference is that $C_{h'}^{\mc{N}=2}$ parameterizes $\mc{N}=2$ descendants: 
\be
C_{h'}^{\mc{N}=2} = d^{\mc{N}=2}_{h'+{k^*\over 2}}~.
\ee
The leading order behavior of the entropy can be read off from (\ref{eq:generalceff}), with $c_{\text{eff}}=3$:
\be
S_{\text{BH}}^{\mc N=2}(0) \approx \pi\sqrt{k^*}.
\ee

In order to get the full perturbative entropy of the $h=0$ black holes (the analog of (\ref{eq:preciselog})), we need to repeat the analyses in Appendix \ref{app:N1Sub} and \ref{app:MultSys} with $\mc{N}=2$ super-Virasoro growth. Unfortunately, for rather technical reasons explained in Appendix \ref{app:MultSys}, it is more difficult in $\mc{N}=2$ theories than in non-supersymmetric or $\mc{N}=1$ theories to obtain the kind of analytic approximation formula for the level densities of the vacuum module that we have been using. However, we can numerically go to fairly high orders in the large $k^*$ expansion.  This in turn gives us a numeric prediction for the entropy of states at $h=0$:
\begin{align}
S_{\text{BH}}^{\mc N=2}(0) \approx \pi\sqrt{k^*} &- \frac{3}{4} \log{k^*} - \log{\(\frac{4\sqrt2}{\pi}\)} - \(\frac{3+\pi^2}{8\pi}\)\frac{1}{\sqrt{k^*}} - \frac{0.03711448}{k^*} \nn \\ &+ \frac{0.21704365}{(k^*)^{3/2}}-\frac{0.13028682}{(k^*)^2}+\mc{O}\(\frac{1}{(k^*)^{5/2}}\).
\label{eq:numericalgarbage}
\end{align}

Just as in $\mc{N}=1$, we have an independent check from positivity in the Ramond sector. We can compute
\be
Z_{\text{R}}(\tau) = \Tr_{\mc{H}_\text{R}}q^{L_0}
\ee
from $Z_{\text{NS}}(\tau)$ by (\ref{eq:modularrns}), and then take $\tau$ to $i \infty$ to find $\text{R}_0$, the number of Ramond sector ground states, which must be nonnegative. We then use (\ref{eq:betadef}) to define an analogous $\beta_{k^*}^{\mc{N}=2}$, and the quantity $(-1)^{k^*+1}\beta_{k^*}^{\mc{N}=2}$ oscillates rapidly between being an upper and a lower bound, making it an independent check for $S_{\text{BH}}^{\mc N=2}(0)$.

In Section 5 of \cite{Gaberdiel:2008xb}, a generating function for $\beta_{k^*}^{\mc{N}=2}$ is derived
\be
\sum_{k^*=1}^{\infty} (-1)^{k^*} \beta_{k^*}^{\mc{N}=2} q^{\frac{k^*}2} = \frac{1-\sqrt{q}}{1+\sqrt{q}} \frac{\theta_3(\tau)}{\eta(\tau)^3}q^{\frac18}(\theta_4(\tau)^4-\theta_2(\tau)^4).
\ee

We have numerically extracted the asymptotic series expansion coefficients of the above expression up to and including ${\cal O}(\frac{1}{k^{*2}})$, and they exactly match (\ref{eq:numericalgarbage}).

\subsection{A No-Go Theorem}
\label{sec:nogo}

The check in the previous subsection was essentially just an extension of the ${\cal N}=1$ case.  However, we will now see that qualitatively new evidence in support of our conjecture is available with ${\cal N}=2$ symmetry when we grade the partition function by the $U(1)$ charge. We will first review a no-go theorem for ${\cal N}=2$ extremal CFTs from \cite{Gaberdiel:2008xb}.   We will then show that our conjecture for the black hole entropies in pure gravity theories precisely evades this no-go theorem.\footnote{The authors of \cite{Gaberdiel:2008xb} also discussed the possible existence of near-extremal CFTs that marginally evaded their theorem, and pointed out that it would be interesting if quantum corrections could be found on the gravity side that would shift pure gravity from an extremal CFT to a near-extremal CFT.  }

Our review of the analysis of extremal $\mc{N}=2$ superconformal field theories in \cite{Gaberdiel:2008xb} will be brief; readers interested in more details should consult the original paper. For simplicity we study chiral, spectral-flow invariant theories with integral $U(1)$ charges and with central charge divisible by $6$. We can define the partition function of a chiral $\mc{N}=2$ SCFT as,
\be
Z(\tau, z) = \Tr_{\mc{H}_{\text{R}}}(-1)^F q^{L_0} y^{J_0}\,, \ \ \ \ \ q\,=\,e^{2\pi i \tau}\,, y=e^{2\pi i z}\,.
\label{eq:n2pf}
\ee
$Z(\tau, z)$ transforms as
\begin{align}
Z\(\frac{a\tau+b}{c\tau+d}, \frac{z}{c\tau+d}\) &= e^{2\pi i m \frac{c z^2}{c\tau+d}} Z(\tau, z) ~~~~~~~~~ \begin{pmatrix} a&b \\ c&d \end{pmatrix} \in SL(2,\mathbb{Z}) \nonumber \\
Z(\tau, z + \ell\tau+\ell') &= e^{-2\pi i m (\ell^2 \tau + 2\ell z)} Z(\tau, z)~~~~~~~~~~~~~~~~~~
\ell,\ell^\prime \in \mathbb{Z}
\label{eq:jacobistuff}
\end{align}
where $m = \frac{c}{6}$.\footnote{In the equation $m=c/6$, $c$ refers to the central charge, not a component of the element of $SL(2,\mathbb{Z})$.} In \cite{Gaberdiel:2008xb}, they considered the more general case of the elliptic genus of any, not necessarily chiral, $\mc{N}=(2,2)$ SCFT; the same analysis follows. We focus on the chiral case, as this relates to our previous discussion.

To make explicit contact with the previous subsection, where we ignored the $U(1)_{\rm R}$, we have,
\es{projz}{
Z_{\rm NS}(\tau)&=q^{m/4}Z\(\tau,\frac{\tau+1}2\)\\
Z_{\rm R}(\tau)&=Z\(\tau,1/2\)\,,
}
where we have used the spectral flow invariance of the $\mathcal{N}=2$ algebra to map between sectors.

An object that transforms as (\ref{eq:jacobistuff}) is known as a weak Jacobi form of weight 0 and index $m = \frac{c}{6}$. See \cite{EichlerZagier} for detailed discussions of Jacobi forms; here we will simply list the pertinent facts used in \cite{Gaberdiel:2008xb}. The ring of weak Jacobi forms is generated by four functions: the Eisenstein series $E_4(\tau)$ and $E_6(\tau)$, as well as the Jacobi forms of weight $0$ index $1$ ($\phi_{0,1}$) and weight $-2$ index $1$ ($\phi_{-2,1}$). These functions are defined in Appendix \ref{app:thetabohnanza}.

A weak Jacobi form of weight $0$ and index $m$ can be written as a linear combination of polynomials
\be
E_4(\tau)^a E_6(\tau)^b \phi_{0,1}(\tau,z)^c \phi_{-2,1}(\tau,z)^d
\ee 
satisfying
\es{eq:linearalgebra}{
4a + 6b - 2d &= 0 \\
c+d&=m.}
At fixed $m$, the dimension of the vector space of weak Jacobi forms of weight 0, index $m$ is simply the number of non-negative integral solutions to (\ref{eq:linearalgebra}). A straightforward count shows that this quantity is given by 
\begin{align}
\text{dim}(V_m) &= \left \lfloor{\frac{m^2}{12} + \frac{m}2 + 1}\right \rfloor \nn \\
&= \frac{m^2}{12}+\frac{m}2 + \mc{O}(1).
\end{align}

The function (\ref{eq:n2pf}) can be expanded as
\be
Z(\tau, z) = \sum_{n, \ell} c(n,\ell) q^n y^\ell.
\label{eq:expand}
\ee
Here, we are counting states labeled by both their dimension and $U(1)_{\rm R}$ charge. The extra quantum number allows for a more refined definition of black hole states. If we define the spectral flow invariant polarity,\footnote{Our definition for polarity differs by a factor of $4m$ from many others.}
\be
P(n,\ell) = n - \frac{\ell^2}{4m}\,,
\ee
then terms in (\ref{eq:expand}) with polarity $P(n,\ell)<0$ are called polar states, and are interpreted as particle states in the dual gravity picture, whereas nonpolar states are the BTZ black holes. Note that some states we would have naively identified as black holes when ignoring the charge are more accurately thought of as particle states.

In \cite{Gaberdiel:2008xb} an extremal $\mc{N}=2$ CFT was defined as a CFT whose partition function \eqr{eq:n2pf}
matches the vacuum character for all polar terms.
However, the number of polar terms goes as $\frac{m^2}{12}+\frac{5m}8$, whereas the dimension of the space of weak Jacobi forms goes as $\frac{m^2}{12}+\frac{m}{2}$. Since the space of polar terms is parametrically larger than the space of weak Jacobi forms, we cannot generically complete an arbitrary polar part into a weak Jacobi form (for instance, by matching the vacuum descendants). Thus at sufficiently large $m$, ${\cal N}=2$ extremal CFTs do not exist.

A natural question to ask, in the spirit of our previous analysis, is how one might relax the criterion of extremality by allowing some states that are not descendants of the vacuum to appear in the theory slightly below or at $n-\frac{\ell^2}{4m}=0$.  This is equivalent to asking how many polar terms can be completed into a weak Jacobi form. In \cite{Gaberdiel:2008xb}, it was shown the number of states with $n - \frac{\ell^2}{4m} < -\frac18$ grows as $\frac{m^2}{12}+\frac{m}2$, which is the same as the dimension of the space of Jacobi forms. So, if we relax our notion of polarity to only include these states, then we generically can complete to a weak Jacobi form, and near-extremal CFTs satisfying this relaxed condition are still viable.

When flowed to the NS sector (without grading by $U(1)$), this notion of near-extremal CFT reduces to our ungraded definition in subsection \ref{sec:ungrade}. Previously, we had defined a near-extremal CFT to be a theory whose partition function matches the vacuum character for all $q^{h}$, with $h<0$. In the ungraded $\mathcal{N}=2$ case a near-extremal theory has an NS sector partition function which matches the vacuum character up to and including the $q^{-1/2}$ term. For the definition of near-extremal given in terms of polarity to be consistent with this earlier definition, it must be the case that modifying the states with polarity $n-\frac{\ell^2}{4m}>-\frac{1}{8}$ does not alter the particle states with $h\leq-1/2$ in the NS sector partition function. To see that the two notions are consistent, note that the polarity is a spectral flow invariant, and so we have $n>-\frac{1}{8}+\frac{\ell^2}{4m}>-\frac{1}{2}$ in the NS sector as well, thus we do not alter the degeneracy of any particle states in the ungraded picture.
\subsection{One-Loop Corrections}
\label{sec:winggkmo}

The relaxed criterion for polarity as $n - \frac{\ell^2}{4m} < -\frac{1}{8}$ looks like a slight downward shift in the threshold energy for black holes in AdS$_3$.  In the case of non-supersymmetric or ${\cal N}=1$ gravity, we found just such a shift; now we want to repeat the analysis for the case ${\cal N}=2$.  This time, because we are grading also by the $U(1)$ charge, we label states by both $L_0$ and $J_0$ eigenvalues. Recall that we can decompose any weak Jacobi form of weight $0$, index $m$ as
\be
Z(\tau,z) = \sum_{r=-m+1}^{m} h_r(\tau) \theta_r(\tau,z)
\ee
with 
\begin{align}
\theta_r(\tau,z) &= \sum_{k=-\infty}^{\infty} q^{\frac{(r+2mk)^2}{4m}}y^{r+2mk} \nn\\
h_r(\tau) &= \sum_{n\geq0}c(n,r)q^{n-\frac{r^2}{4m}}
\end{align}
and $c(n,r)$ defined in (\ref{eq:expand}). Moreover, $h_r(\tau)$ is a vector-valued modular form of weight $-\frac12$. The subscript $r$ refers to the sector the state is in, which is the $U(1)$ charge mod $2m$. From e.g. \cite{Dijkgraaf:2000fq, Manschot:2007ha, Whalen:2014}, we can write the coefficient of an arbitrary term $q^\D$, $\D>0$, in $h_{\nu}$ as a vector-valued Rademacher sum. We can organize this as an infinite sum over all elements of $SL(2,\bb{Z})$ (modded out by $\tau \rightarrow \tau+n$), but in the large $m$ limit, the dominant contribution comes from the image of $\tau \rightarrow -\frac1\tau$. Keeping only this gives
\be
F_{\nu}(\D) \sim (2\pi)^{5/2} \sum_{\mu=1}^{2m} e^{-\frac{i\pi\mu\nu}{m}} \sum_{\alpha<0} |\alpha|^{3/2} F_{\mu}(\alpha) I_{\frac32}(4\pi \sqrt{|\alpha| \D}).
\label{eq:vv}
\ee
In this formula, the parameters $F_\mu(\alpha)$ are the number of polar states in sector $\mu$ and polarity $\alpha<0$, i.e. 
\be
F_{\mu}(\alpha) \equiv c\(\frac{\mu^2}{4m} + \alpha,\mu\).
\label{eq:polarbear}
\ee
The output $F_{\nu}(\D)$ is the number of states in sector $\nu$ with polarity $\D>0$. Our goal is then is to use (\ref{eq:vv}) to calculate the entropy of a state in sector $\nu$ and polarity $\D$ (which can be converted to the standard quantum numbers using (\ref{eq:polarbear})). Just as in previous sections, we will do this by finding the dominant contribution in the double sum in (\ref{eq:vv}) where there is a tradeoff between the Bessel function and the growth of polar states.

For an extremal $\mc{N}=2$ theory, polar states are given by the number of descendants in the Ramond sector. The generating function for Ramond sector vacuum descendants\footnote{Here by Ramond sector vacuum descendants, we mean descendants of the state that is the spectral flow image of the unique NS vacuum, in other words descendants of the charge $m$ Ramond vacuum.} is
\begin{align}
\sum_{n,\ell} d\(n,\ell\)q^{n}y^\ell &= (1-q)y^m \prod_{n=1}^{\infty} \frac{(1-yq^{n+1})(1-y^{-1}q^n)}{(1-q^n)^2}\nn \\
&= \frac{(1-q)y^{m-\half}}{(1-y^{-1})(1-yq)} \frac{i\theta_1(\tau,z)}{\eta(\tau)^3}
\label{eq:rgs}
\end{align}
where $d(n,\ell)$ is the number of vacuum descendants with quantum numbers $L_0=n, J_0=\ell$.
To estimate the growth of the polar state degeneracies, we have to estimate the growth of the RHS of (\ref{eq:rgs}). We see numerically that
\be
d\(n,\ell+m-\half\)=(-1)^{\ell-\half} e^{\pi \sqrt{2 \(n-\frac{\ell^2}2\)}}\(\frac{64}{\pi\(n-\frac{\ell^2}2\)^2} + \mc{O}\(n-\frac{\ell^2}2\)^{-5/2}\).
\label{eq:fitsn2}
\ee
Note that up to a phase of $(-1)^{\ell-\half}$, (\ref{eq:fitsn2}) is a function only of $n-\frac{\ell^2}2$, the polarity of a theory at $c=3$. Indeed, the number of descendants of the $\mc{N}=2$ SCFT vacuum should grow roughly as a CFT with $c=3$. Moreover the phase in (\ref{eq:fitsn2}) will not affect the leading order behavior of the entropy.\footnote{This is because $\ell$ can only take half-integral values (since we defined the $U(1)$ charge as $\ell+m-\half$), so half of all vacuum descendants will contribute with positive sign, and half will contribute with negative. The positive and negative contributions are not the same perturbatively, so we do not have full cancellations between the two and the leading order stays the same.}

Using (\ref{eq:fitsn2}) we have an estimate of $F_\mu(\a)$ in (\ref{eq:vv}); the remaining term that is exponentially large is the Bessel function $I_{\frac32}(4\pi\sqrt{|\a|\D})$. Since we are considering the effect of a state $q^n y^{\ell+m-\half}$, the polarity $\a$ is given by
\begin{align}
\a &= n - \frac{(\ell+m-\half)^2}{4m} \nn \\
&= - \frac m4 + n - \frac\ell2 + \frac14 + \mc{O}\(\frac1m\)
\end{align}
and thus
\be
I_{\frac32}(4\pi\sqrt{|\a|\D}) \sim e^{4\pi \sqrt{\D \(\frac{m}4 + \frac{\ell}{2} - \frac14 - n\)}}.
\ee

We see the dominant contribution comes from
\begin{align}
\ell^* &= \half \nn \\
n^* &= \frac{m+4\D}{32\D+4}
\end{align}
which gives
\be
S^{\mc N=2}(\Delta, \nu) = \half \pi \sqrt{(2m-1)(8\D+1)}
\label{eq:pre18shift}
\ee
or
\be
S^{\mc N=2}(n, \ell) = 2\pi \sqrt{\(m-\frac12\)\(n - \frac{\ell^2}{4m} + \frac18\)}
\label{eq:18shift}
\ee
as the perturbative piece.

Interpreting (\ref{eq:18shift}) as before, this suggests that states with polarity
\be
n-\frac{\ell^2}{4m}>-\frac18
\label{eq:barely}
\ee
may be regarded as black holes. Thus, we have confirmed the prediction of \cite{Gaberdiel:2008xb}, based on the difference in the space of Jacobi forms and polar parts, by direct calculation.

It would be interesting to find corrections to (\ref{eq:barely}) that match the higher order discrepancies between the space of Jacobi forms and polar parts. Unfortunately this is difficult for two reasons. Firstly, the next order difference between the two grows as $\mc{O}(m^{\half})$, but does not converge to an actual coefficient; instead it behaves roughly as a random variable. See \cite{Gaberdiel:2008xb, EichlerZagier} for more discussion on this point. Secondly, it is not clear how to proceed finding corrections to (\ref{eq:barely}). It is not done by finding higher order corrections to (\ref{eq:18shift}) -- that would instead correct the entropy of those black holes. Instead, we would like to find corrections inside the square root in (\ref{eq:18shift}), so that we have more discrimination in finding which quantum numbers lead to nonnegative entropy.

\section{Towards a Non-Perturbative Conjecture}\label{secva}

A natural question that remains is whether one can find a non-perturbative completion of the perturbative conjecture for $N_{\rm BH}(0)$, the number of states at the black hole threshold. Such a completion would be indispensable in trying to construct explicit models of CFTs dual to pure gravity for any $k$. Here we discuss two approaches to this question. The first is a very natural, but ultimately incomplete, attempt at a non-perturbative completion via direct analytic continuation of the Rademacher sum to $h=0$. We do this for ${\cal N}=0,1$. The second, which is more speculative, is a construction of a ``grand canonical'' partition function in the space of chiral CFTs, whose modular properties are directly related to the number of threshold states. Though this object {\it a priori} contains no new information, we will see that its form suggests an intriguing larger symmetry. 
We are hopeful that our attempts will inspire further investigations. 

\subsection{Analytic Continuation of Rademacher Sums}

The goal of this subsection is to propose a non-perturbative formula for the number of states at threshold, $N_{\rm BH}(0)$, in non-supersymmetric chiral gravity, extending (\ref{eq:N0log}). (We will treat the ${\cal N}=1$ case next.) For states with positive dimension above threshold, a non-perturbative formula for the degeneracy is given by a Rademacher sum for the coefficients of modular forms. As explained in Appendix \ref{app:MultSys} the number of states at level $h$ is given by,
\es{radfullsum}{
N_{\rm BH}(h)&= 2\pi \sum_{h'=-k}^{-1} C_{h'} \sqrt{\frac{-h'}{h}} \sum_{n=1}^\infty \frac{1}{n} A_{n,-h'}(h) I_1\(\frac{4 \pi}{n} \sqrt{-h h'}\)\,.
}
$A_{n,-h'}(h)$ is known as a Kloosterman sum. This expression admits a continuation to $h=0$.\footnote{This analytic continuation has been considered in the math literature before: our expression for $N_{\rm BH}^{NP}(0)$ is known there as the ``Rademacher constant.'' See e.g. Section 5.1 of \cite{Duncan:2009sq}, or \cite{Cheng:2012qc}.} The Bessel function gives
\eq{}{I_1\(\frac{4 \pi}{n} \sqrt{-h h'}\)\approx {2\pi\over n}\sqrt{-h h'}+\ldots}
As we review in Appendix \ref{app:MultSys}, the Kloosterman sum at $h=0$ reduces to a Ramanujan sum; the resulting sum over $n$ yields
\eq{}{\sum_{n=1}^{\infty}{A_{n,-h'}(0)\over n^2}  = {24\over (2\pi)^2}{\sigma_1(-h')\over -h'}}
where $\sigma_1(-h')$ is the divisor function. Putting things together, we find
\es{npguesssum0}{
N_{\rm BH}^{NP}(0)&= 24 \sum_{h'=-k}^{-1} C_{h'} \sigma_1(-h')\,.
}
The superscript denotes ``non-perturbative.'' This all goes through for any light spectrum $C_{h'<0}$; to specialize to pure gravity, recall that the number of polar states $C_{h'<0}$ can be written in terms of the integer partition function $p(n)$,
\es{n0desc2}{
C_{h'}&= p(h'+k)-p(h'+k-1)\,.
}
The sum (\ref{npguesssum0}) can actually be done in closed form using \eqr{n0desc2} and a basic property of $\sigma_1$, 
\eq{conv}{\sum_{n=1}^k \sigma_1(n) p(k-n) = k p(k)}
giving
\es{npgues0}{
N_{\rm BH}^{NP}(0)&=24\left(k p(k)-(k-1)p(k-1)\right)\,.
}

This gives the following extremal partition functions for low values of $k$: 
\es{}{
Z_1 &= q^{-1} + 24 + 196884\,q+ \ldots \nn \\
Z_2 &= q^{-2} + 72 + 42987520\,q + \ldots \nn \\
Z_3 &= q^{-3}+q^{-1} + 120 + 2593096794\,q + \ldots \nn \\
Z_4 &= q^{-4}+q^{-2}+q^{-1} + 264 + 81026609428\,q + \ldots. \nn} 

This appealing result passes a couple of non-trivial checks. Firstly, the proposed constant for each partition function is a positive integer. This is a necessary condition that any sensible conjecture must satisfy. Secondly, the chiral CFTs at $k=1$ have been classified by Schellekens \cite{Schellekens:1992db}. A necessary condition for any non-perturbative guess is that the number of currents for the $k=1$ theory match one of the allowed values in Schellekens' classification. The result $N_{\rm BH}(0)=24$ for $k=1$ does indeed fall into this list: namely, it matches the theory of $24$ bosons compactified on the Leech lattice. This is especially interesting because the Leech lattice CFT is the unorbifolded parent of the Monster CFT, which is Witten's extremal CFT at $k=1$.

Before we can claim success, however, we must check whether the perturbative expansion of \eqr{npgues0} at large $k$ matches with the perturbative result (\ref{eq:N0log}) computed earlier. Expanding both, the non-perturbative conjecture (\ref{npgues0}) gives
\es{nonpertexp}{
N_{\rm BH}^{NP}(0)&=\frac{48 \sqrt{3}}{\pi} \left(\frac{e^{\frac{1}{6} \pi  \sqrt{24 k-1}} k \left(\pi  \sqrt{24 k-1}-6\right)}{(24 k-1)^{3/2}}-\frac{e^{\frac{1}{6} \pi  \sqrt{24 k-25}} (k-1) \left(\pi  \sqrt{24 k-25}-6\right)}{(24 k-25)^{3/2}}\right)+\ldots\\
S_{\rm BH}^{NP}(0)&=\log\left(N_{\rm BH}^{NP}(0)\right)\,=\,\pi\sqrt{\frac{2}{3}k}+\frac{1}{2}\log\left(\frac{2\pi^{2}}{k}\right)-\frac{72+13 \pi ^2 }{24 \sqrt{6} \pi }\sqrt{\frac{1}{k}}+\ldots
}
while the perturbative result (\ref{eq:N0log}) is
\es{pertexp}{
S_{\rm BH}(0)&=\pi\sqrt{\frac{2}{3}k}+\frac{1}{2}\log\left(\frac{2\pi^{2}}{k}\right)-\frac{13 \pi ^2 }{24 \sqrt{6} \pi }\sqrt{\frac{1}{k}}+\ldots
}
Unfortunately these differ at order $1/\sqrt{k}$. This tells us that, at least at large $k$, the non-perturbative conjecture \eqr{npgues0} is incomplete.\footnote{Conceivably, \eqr{npgues0} might nevertheless be correct for low values of $k$, where the perturbative expression is no longer even approximately expected to be valid.  However, we are reluctant to trust an expression for the exact number of states at low $k$ when we have presented evidence that the same expression should not be correct at large values of $k$.} 

\subsubsection*{$\mathcal{N}=1$}
The logic of the previous subsection carries over almost unchanged to the supersymmetric case. Analytically continuing the ${\cal N}=1$ Rademacher sum of Appendix \ref{app:MultSys} to $h=0$ for an arbitrary light spectrum $C^{{\cal N}=1}_{-\frac{h'}2}$  gives the conjectural number of threshold states to be
\es{eq:nonpertn1}{
N^{NP}_{\text{BH}}(0)=\sum_{h'=1}^{k^*} 8f(h') C^{{\cal N}=1}_{-\frac{h'}2}
}
where 
\eq{}{f(h')=|(\text{sum of even divisors of~}h')-(\text{sum of odd divisors of~}h')|}
In pure ${\cal N}=1$ supergravity, $C^{{\cal N}=1}_{n} = d_{n+{k^*\over 2}}^{{\cal N}=1}$, cf. \eqr{supervac}. This gives the following NS sector partition functions for low values of $k^*$:
\begin{align}
Z_1^{\mc{N}=1}&= q^{-1/2} + 8 + 276\,q^{1/2} + \ldots \nn \\
Z_2^{\mc{N}=1}&= q^{-1}+ 8 + 4096\,q^{1/2} + \ldots \nn \\
Z_3^{\mc{N}=1}&= q^{-3/2} + 32 + 33606\,q^{1/2}+ \ldots \nn \\
Z_4^{\mc{N}=1}&= q^{-2}+q^{-1/2}+40+196884\,q^{1/2}+\ldots.  
\end{align}

Again, it is encouraging that \eqr{eq:nonpertn1} predicts a non-negative integer number of states at threshold. Furthermore, the case $k^{*}=1$ corresponds to a known theory: $Z_1^{\mc{N}=1}$ is exactly the partition function for 8 free bosons compactified on the $E_8$ root lattice, along with their superpartners. Orbifolding this $E_8$ theory by the natural $\mathbb{Z}_2$ gives the Conway CFT, in the same way that orbifolding the Leech lattice gives the Monster CFT.

\begin{figure}[t!]
   \centering
    \includegraphics[width=0.75\textwidth]{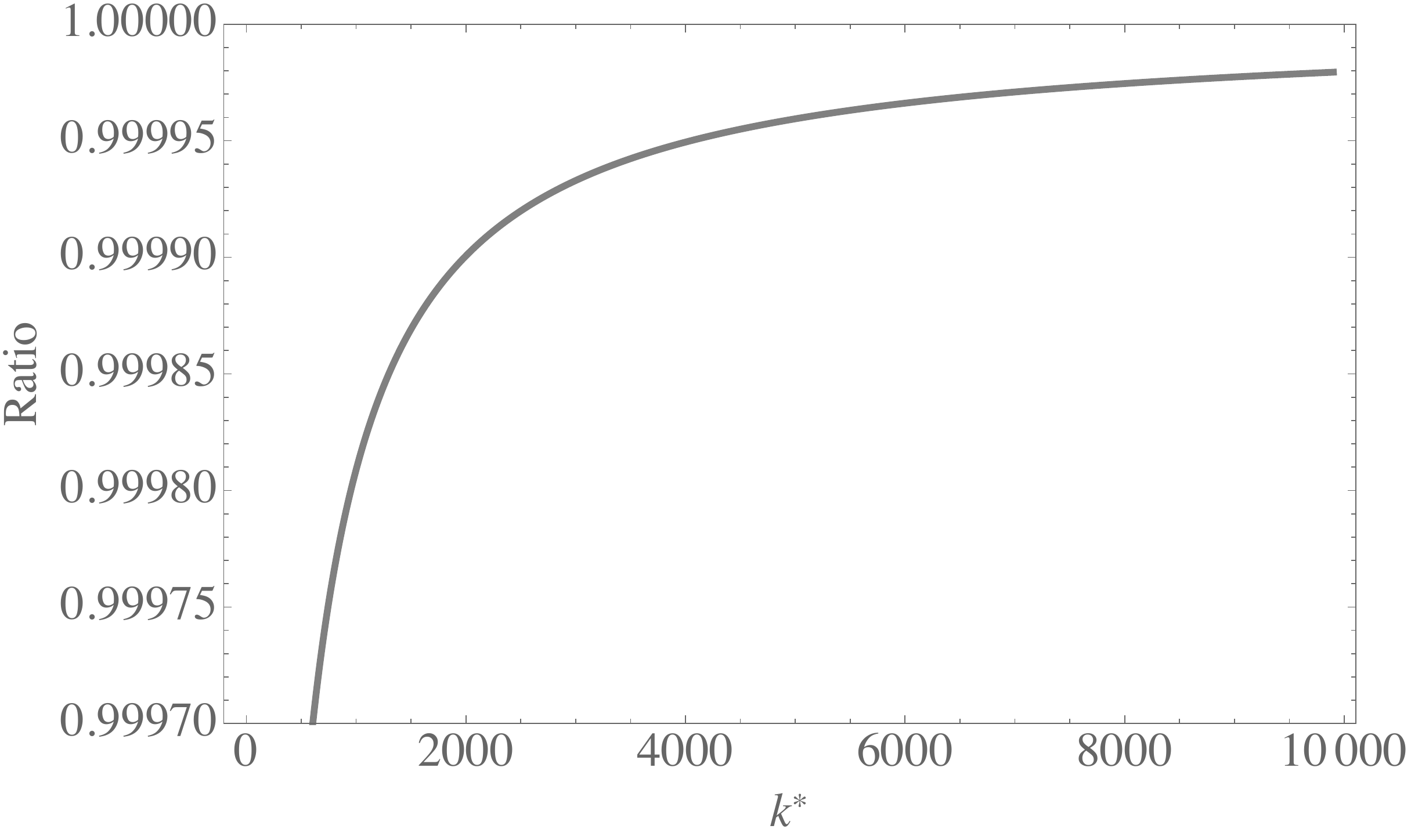}
    \caption{Here we plot the ratio of the log of the number of NS primaries at $\frac{k^*}{2}$ predicted by our analytic continuation \eqr{eq:nonpertn1} to $\log |\beta_{k^*}|$, as a function of $k^*$. Although these functions are extremely close (differing by 0.002\% at $k^*=10000$), the number of predicted $h=0$ NS primaries is always less than $|\beta_{k^*}|$. This means that for even $k^*$, where $|\beta_{k^*}|$ is a lower bound for the number of primaries, the prediction is not consistent with unitarity.}
 \label{fig:nonpertbeta}
\end{figure}

However, unitarity also requires that any conjecture must predict a non-negative number of states in the Ramond sector as well. Unfortunately, this check fails starting at $k^*=16$, and for all even values of $k^*\ge 16$ this conjecture (barely) predicts a negative number of Ramond ground states; see Figure \ref{fig:nonpertbeta}. So while this non-perturbative guess comes remarkably close to predicting a viable partition function,  it ultimately conflicts with unitarity, at least for even $k^* \ge 16$. 

While a logical possibility is that \eqr{eq:nonpertn1} is correct for all odd $k^*$, this is obviously not completely satisfactory. Similarly, \eqr{eq:nonpertn1} may also be correct for $k^*<16$ only. We believe the story at both ${\cal N}=0$ and ${\cal N}=1$ to be tantalizing, but incomplete; we hope that these conjectures may ultimately be augmented or better understood. 

\subsection{A Modular Approach}

In general, given an infinite family of CFTs indexed by some integer $N$, one can construct a generalized partition function by summing over the partition functions $Z_N(\tau)$, weighted by a fugacity whose chemical potential is conjugate to $N$. We carry this out below for chiral CFTs, indexed by $k=c/24$, whose polar spectrum takes a fixed structure.
Some comments related to this will appear in \cite{MdLV}.

Suppose we have a sequence of chiral partition functions $Z_{k}(\tau)$ for theories with central charge $c=24k$. We can construct the generating function
\es{ceq}{
\mathcal{Z}(\sigma,\tau)&=\sum_{k}p^{k}Z_{k}(\tau)\,.
}
with $p\equiv e^{2\pi i \sigma}$, $q\equiv e^{2\pi i \tau}$.
Now suppose the polar spectrum is given by 
\es{lightz}{
Z_{k}(\tau)=q^{-k}(1+c_{1}q+c_{2}q^{2}+\ldots+c_{k-1}q^{k-1})+\text{non-polar}
}
where the $c_i$ are the same for all $k$; only the location of the threshold changes. Let's denote the generating function of this fixed light spectrum as, 
\es{gf}{
\sum_{i=0}^{\infty}c_{i}q^{i}&\equiv\mathcal{X}(q)\,,
}
where $c_{0}\equiv 1$. Now we can begin to construct the generating function,
\es{pertgc}{
\mathcal{Z}(\sigma,\tau)&=p\left(\frac{1}{q}+\ldots\right)+p^{2}\left(\frac{1}{q^{2}}+c_{1}\frac{1}{q}+\ldots\right)+\ldots\\
&=\frac{1}{q}\sum_{i=0}^{\infty}c_{i}p^{i+1}+\frac{1}{q^{2}}\sum_{i=0}^{\infty}c_{i}p^{i+2}+\ldots\\
&=\mathcal{X}(p)\sum_{i=1}^{\infty}(p/q)^{i}+\ldots\\
&=\frac{p\mathcal{X}(p)}{q-p}+\ldots\,,
}
Here the ellipses represents our ignorance about the non-polar terms, but these can be fixed by using modular invariance. In particular, we want a modular invariant function of $q$ that has a simple pole at $p$ with the correct residue. This fixes $\mathcal{Z}(\sigma,\tau)$ up to an overall function of $p$, call it $R(p)$, which encodes our ignorance about the number of states at threshold:
\es{Full}{
\mathcal{Z}(\sigma,\tau)&=\frac{\partial_\sigma j(\sigma)}{2\pi i(j(\tau)-j(\sigma))}\mathcal{X}(p)+R(p)\\
&=\frac{E_{4}^{2}(\sigma)E_{6}(\sigma)}{\Delta(\sigma)(j(\sigma)-j(\tau))}\mathcal{X}(p)+R(p)\,
}
where we recall the definition $\Delta(\sigma) = \eta^{24}(\sigma)$. For the case of extremal or near-extremal CFTs, the generating function $\mathcal{X}(p)$ is just the Virasoro vacuum character (times $q^k$),
\es{dgen}{
\mathcal{X}(p)&=(1-p)\frac{p^{1/24}}{\eta(\sigma)}\,,
}
and the grand canonical partition function is,
\es{gcext}{
\mathcal{Z}(\sigma,\tau)&=(1-p)\frac{E_{4}^{2}(\sigma)E_{6}(\sigma)}{\Delta(\sigma)(j(\sigma)-j(\tau))}\frac{p^{1/24}}{\eta(\sigma)}+R(p)\,.
}
If we take the number of states at $h=0$ to be \eqr{npgues0}, we get
\eq{rofp}{R(p) = -{(1-p)}\frac{p^{1/24}}{\eta(\sigma)}{E_2(\sigma)}\,,}
where $E_2(\sigma)$ is the second holomorphic Eisenstein series, defined in Appendix \ref{app:thetabohnanza}; one can check this by expanding \eqr{gcext} at small $p$ for fixed $q$, and comparing to $\sum_k p^k N_{\rm BH}^{NP}(0)$. Perhaps suggestively, this has nice modular properties. We note for comparison that a grand canonical sum over extremal CFT partition functions gives $R(p)=0$.

Without any additional input, we do not know what modular properties the grand canonical partition function should have; if we did, this would fix $R(p)$, thereby giving an answer for the number of black hole states at threshold. However, the form of $\mathcal{Z}(\sigma,\tau)$ suggests that perhaps the correct answer has some well behaved modular behavior in $\sigma$. This would be a very exciting way to access quantum gravity in the Planckian regime.
Some comments  on modular transformation properties of grand canonical partition functions will appear in \cite{MdLV}.

\section{Small Black Holes From Modularity in Chiral CFT}\label{secv}
Implicit in our discussion so far has been an interpretation in terms of a holographically dual chiral (or holomorphically factorized) CFT. To summarize, we have argued that the dual to pure gravity is not an extremal CFT, but rather a near-extremal CFT which has $N_{\rm BH}(0)$ states at level $k$, where $N_{\rm BH}(0)$ was given to all orders in $1/k$ perturbation theory in \eqr{eq:N0log}. Analogous statements hold for pure ${\cal N}=1,2$ supergravities. 

Given that we lack a complete conjecture for $N_{\rm BH}(0)$ valid at all $k$, a construction of these putative near-extremal CFTs at low $k$ is not yet possible. However, chiral CFTs are subject to strong constraints imposed by modularity, holomorphy and crossing symmetry, which help define the landscape of chiral CFTs. Some basic facts about chiral CFTs are reviewed in Appendix \ref{chiapp}. In this section, we derive one such result that is especially relevant to the preceding gravity computations: in any chiral CFT with spin-1 currents, the number of primary states at level $k$ is bounded from below. Moreover, at large $k$, this lower bound grows exponentially with $\sqrt k$, mimicking the behavior of $N_{\rm BH}(0)$. 

Our starting point is the partition function, 
\eq{ztau}{Z(\tau) = \Tr q^{L_0} = \sum_{h= -k}^{\infty} C_hq^{h},}
where $h\in \Z$ in a chiral CFT. 
We may also decompose $Z(\tau)$ into Virasoro characters,
\eq{}{Z(\tau) = \sum_{h= -k}^{\infty} N_{h+k}\,\chi_{h}(\tau),}
We have introduced a new notation for later convenience: 
\es{}{C_{-k+\ell} &= \text{total number of states at level $\ell$ above the vacuum},\\
N_{\ell} &= \text{total number of Virasoro primary states at level $\ell$ above the vacuum}.}
A physical CFT has non-negative integer degeneracies: $N_{\ell}\in \mathbb{N}$. 

As we review in Appendix \ref{chiapp}, $Z(\tau)$ is a meromorphic function on the Riemann sphere $\C^*$ with a $k^{\text{th}}$ order pole at $\tau=i\infty$ and no other poles. Accordingly, it can be written as 
\eq{}{Z(\tau) = \Delta(\tau)^{-k} {\cal Z} (\tau), }
where $\Delta(\tau)=\eta^{24}(\tau)$ is the modular discriminant, which is a modular form of weight $12$, and ${\cal Z} (\tau)$ is a modular form of weight $12k$. Similar statements hold for $n$-point functions; while for $n>1$ there are modular ``anomalies,'' one can still derive powerful conclusions using modularity. 

To demonstrate this, we now consider a chiral CFT with spin-1 currents $J^a$, where $a=1,\dots, N_1$, and perhaps other polar primaries too.  Rather than computing the partition function, we can construct the weighted partition function,
\be
P(\tau, F^a ) = \Tr e^{F^a J^a_0} q^{L_0},
\ee
where we have introduced a potential $F^a$ for each current.  This function is not modular invariant, but its transformation properties are known \cite{Schellekens:1992db}. The simplest way to obtain interesting constraints is to expand in powers of $F$.  We have 
\be
P(\tau, F^a )= Z(\tau) + P_{aa}(\tau) F^{a}{}^2 + \dots,
\ee
where 
\eq{}{P_{aa}(\tau) \equiv \half \Tr(J_0^aJ_0^a q^{L_0})\geq 0}
is the current two-point function on the torus. Importantly, this must be non-negative. We can constrain the form of this object using modularity. First, at the cusp at infinity, $P_{aa}(\tau=i\infty)\sim q^{-k}$, coming from the identity piece of the trace. $P_{aa}(\tau)$ almost transforms with modular weight two, but for an ``anomaly'' factor that may be computed explicitly \cite{Schellekens:1992db}. Altogether, one may write $P_{aa}(\tau)$ as
\be\label{paa}
P_{aa}(\tau) =  \Delta(\tau)^{-k} \eps_{12k+2} +{1\over 48} E_2(\tau) Z(\tau) ,
\ee
where $\eps_{12k+2}$ is an $SL(2,\Z)$ modular form of weight $12k+2$. 

We now note the following very useful fact: $\Delta^{-k} \epsilon_{12k+2}$ has a vanishing constant piece for any modular form $\epsilon_{12k+2}$! The proof is simple. Define the modular derivative
\eq{0c}{D_w \equiv q d_q -{w\over 12} E_2~.}
Its action on weight-$w$ forms takes them to weight-$(w+2)$ forms. Using 
\eq{}{q \p_q \log\D = E_2,}
we can generate any weight $12k+2$ modular form $\eps_{12k+2}$ in terms of a weight $12k$ modular form $\eps_{12k}$ as\footnote{The space $M_{12k}$ of weight-$12k$ modular forms is dimension $k+1$, whereas the space $M_{12k+2}$ of weight-$12k+2$ modular forms is dimension $k$, so  $D_{12k}$ annihilates the weight-$12k$ modular form $\Delta^k$ and takes the rest of $M_{12k}$ onto $M_{12k+2}$.}
\eq{1i}{\D^{-k} \eps_{12k+2} = \D^{-k} D_{12k}\eps_{12k}
= qd_q(\D^{-k} \eps_{12k}),}
which concludes the proof.

This fact leads to interesting constraints on the constant term of \eqr{paa}: in particular, the contribution of $\Delta^{-k}\eps_{12k+2}$ vanishes. Given that the correlator $P_{aa}(\tau)$  must have a $q$-expansion with non-negative coefficients, imposing this condition relates the degeneracies of the ``light'' operators with $-k\leq h<0$ to one another, due to the appearance of $Z(\tau)$ in \eqr{paa}. By demanding positivity of the constant term of $P_{aa}(\tau)$, one finds the following: for $any$ chiral CFT with spin-1 currents, the total number of states at $h=0$, namely $C_0$, is bounded as
\be\label{bound}
C_0 \ge 24 \sum_{h=-k}^{-1}  C_{h} \sigma_1(-h)~.
\ee
This follows simply from \eqr{ztau} and \eqr{e2def}.

For example, in a $k=2$ chiral CFT with $N_1$ currents, positivity of $P_{aa}$ means that the total number of spin-2 states, $C_{0}$, is bounded as
\be
C_{0} \ge 24 (N_1+3).
\ee
Let us consider this constraint in the case of $U(1)$ currents. If we assume that the stress tensor is independent of the $U(1)$ currents, there are $1+N_1(N_1+3)/2$ descendants under the Virasoro and current algebra at level 2. So the number of spin-2 primaries, $N_2$, is bounded as
\be
N_{2} \ge - {1\over 2} N_1^2 + {45\over 2} N_1 +71~.
\ee
This implies that there must exist spin-2 primaries whenever $1\leq N_1 < 48$. In fact, if the stress tensor is independent from the descendants of the $U(1)$ currents, $N_1\leq 48$ is already required in order not to violate Cardy's formula. If we assume a single null vector at level 2 (which would be the case if we have a Sugawara stress tensor) then our inequality becomes trivial, $N_2\geq 0$, when $N_1=48$. This would be saturated for free bosons on a 48-dimensional even self-dual lattice.

\subsection*{Large k}
At large $k$, we can say more: after subtracting all descendants at $h=0$, the bound \eqr{bound} translates to a lower bound on the number of level-$k$ {\it primaries}, $N_k$, which grows exponentially in $\sqrt k$ with a positive overall coefficient. This is evident from \eqr{bound}, as we now explain. First, it's clear that the number of descendants at level $k$ is less than the right-hand side of \eqr{bound}, which is the sum over the {\it total} number of states at all levels $\ell<k$, weighted by $24 \sigma_1(k-\ell)\geq 24$. This proves that there exists a non-trivial bound on $N_k$. To see that it grows exponentially with $\sqrt k$, we approximate $C_h$ by its Cardy growth
\eq{casy}{C_h \approx e^{2\pi\sqrt{c_{\rm eff}({(h+k)\over 6}-{c_{\rm eff}\over 24})}}}
where we ignore subleading asymptotics. $c_{\rm eff}$ counts the number of light primaries in the CFT; for a theory of $N_1$ spin-1 currents and the Virasoro vacuum module, for instance, $c_{\rm eff}=N_1+1$. Then subtracting the descendants, which scale the same way as \eqr{casy}, gives
\eq{}{N_k \ge 24 \left(\sum_{h=-k}^{-1}  e^{2\pi\sqrt{c_{\rm eff}({(h+k)\over 6}-{c_{\rm eff}\over 24})}}\sigma_1(-h)\right) - e^{2\pi\sqrt{c_{\rm eff}({k\over 6}-{c_{\rm eff}\over 24})}}~.}
Given that $\sigma_1(k)$ grows sub-exponentially for large $k$,
\eq{}{\limsup_{k\rightarrow\infty}\,{\sigma_1(k) \over k \log \log k} = e^{\gamma}}
where $\gamma$ is the Euler-Mascheroni constant (e.g. \cite{Hardy}, p.266), we arrive at 
\eq{nkbound}{N_k\gtrsim  e^{2\pi\sqrt{{c_{\rm eff}k\over 6}}}}
to leading order in large $k$.

This exponential lower bound has the same scaling of $N_{\rm BH}(0)$ in pure gravity derived in Section \ref{sec:N0}, albeit with a different value of $c_{\rm eff}$ reflecting the presence of currents. Even by adding a single $U(1)$ current to the gap, we have shown, using an independent chiral CFT argument, that the number of  threshold primaries $N_k$ at level $k$ is bounded below by an exponentially growing function of $\sqrt k$.  An important caveat is that the presence of a current in the theory also changes the threshold for black holes and makes it charge-dependent,\footnote{This is also clear from the ${\cal N}=2$ discussion in Section \ref{sec:n2}, and from the definition of the black hole threshold for bosonic charged BTZ black holes, where the spectral flow invariant sets the black hole threshold at leading order in large $k$.} e.g.  in AdS$_3$ gravity plus a gauge field a  horizon forms when $m_{\rm BH} \gtrsim \frac{1}{8 G_{\rm N}} + \frac{Q^2}{2}$ for an abelian current with central charge 1.  Therefore,  not all of the $N_k$ threshold primaries need be dual to black holes: in particular, some of them may sit slightly below the black hole threshold if they have sufficiently large charge. Nevertheless, if there are ``pure gauge plus gravity'' theories that do not have such states, (\ref{nkbound}) applied to them would be a lower bound on the number of neutral black hole states at threshold.

In all, this bound on $N_k$ suggests a universality of our gravity result even in the presence of matter and, conversely, gives CFT support to the bulk computation. An obvious goal would be to determine, from CFT, whether \eqr{nkbound} holds when continued to the pure gravity value $c_{\rm eff}=1$. If so, this would rule out extremal CFTs at large $k$. Our approach here is similar to adding supersymmetry, and is indicative of a more general lesson: given any spin-$s$ current, the graded partition function is subject to new modular constraints. We hope to report more on this ``chiral bootstrap'' in the future.

\section{Discussion}\label{secvi}

In this paper, we have computed quantum corrections to the entropy of small black holes in AdS$_{3}$ pure gravity. As a result we were able to make a conjecture for the entropy of black holes with vanishing classical horizon area. In the supersymmetric context, this conjecture passed an all-orders consistency requirement. Furthermore, as we have emphasized, this increased number of states at threshold allows the $\mathcal{N}=2$ extremal theories to evade the no-go theorem of \cite{Gaberdiel:2008xb}. These computations in extremal and near-extremal gravity theories represent one of the cleanest possible contexts in which to study quantum corrections to black hole entropy, as well as the thermodynamic properties of small black holes.

Though our computations were focused on the case of pure gravity, the results have applicability even with additional matter. If we allow extra light states, as in our CFT computation of Section \ref{secv}, our formula for the degeneracy of light black holes (\ref{eq:nokloos}) gets modified to
\es{moregen}{
N_{\text{BH}}(h)\propto e^{4\pi\sqrt{\(k-\frac{1}{24}\)\(h+\frac{1}{24}\)}}+\sum_{-k<h^{\prime}<0}N_{h^{\prime}+k}e^{4\pi\sqrt{\(|h^{\prime}|-\frac1{24}\)\(h+\frac1{24}\)}}\,,
}
where the second term now incorporates contributions from light states above the vacuum, and $N_{h^{\prime}+k}$ is the number of primary states at level $h'+k$ above the vacuum. For sufficiently sparse theories the leading order black hole entropy is unchanged from our result (\ref{eq:doitright}).
\es{sbhlead}{
S_{\text{BH}}(h)\sim4\pi\sqrt{\left(k-\frac{1}{24}\right)\left(h+\frac{1}{24}\right)}\,.
}
Furthermore, the contributions from additional light states when naively continued to $h=0$ serve to increase the number of black holes at threshold.  

There are a number of unanswered questions, however. Chief among them is whether fully consistent near-extremal CFTs exist. For the case of extremal CFT there have been a variety of consistency checks, \cite{Witten:2007kt,Gaiotto:2007xh,Yin:2007gv,Gaberdiel:2010jf} and no-go theorems \cite{Gaberdiel:2007ve,Maloney:2007ud,Gaiotto:2008jt,Gaberdiel:2008xb}. For near-extremal theories even less is known. We have advocated that pure gravity requires a minimal version of non-extremality, modifying only the number of threshold black hole states. Our conjecture for the degeneracy of these states introduces a strong, novel constraint, and will hopefully be useful in fully determining the dual CFTs. A tantalizing intermediate step is the question of the correct non-perturbative completion of our conjecture for the number of states at $h=0$.
While the non-perturbative completions we considered in Section \ref{secva} turned out not to match our perturbative results of Sections \ref{sec:N0} and \ref{sec:n1} when expanded at large central charge, they were quite close. This raises an interesting possibility, not explored in detail here, that allowing a relatively small number of additional states below $h=0$ may allow for such match.

Another interesting direction is how to understand the quantum corrections to black hole entropy from a spacetime calculation. For instance, is it possible to identify the saddles contributing the entropy of the black holes with zero classical size?

From the more algebraic side, it is clear that modularity of the torus and higher genus partition functions, particularly in conjunction with crossing symmetry of four-point functions, provides especially strong constraints in sparse chiral CFTs \cite{Witten:2007kt,Gaiotto:2007xh}. It would be fantastic if these constraints could be harnessed to complete the construction of some subclass of holographic CFTs, or to further narrow the landscape of such theories \`a la the conformal bootstrap. 

To conclude, while chiral theories lack some of the hallmarks of more familiar gravity, such as low spin black holes, bulk local degrees of freedom, and chaos, they do resemble the gravitational sector of traditional non-chiral theories in AdS$_3$, and can be thought of as a laboratory for studying gravitational interactions between non-chiral degrees of freedom. Given the powerful techniques available in AdS$_3$/CFT$_2$, we expect that further results along the lines of what we have presented here are in the offing. 

\section*{Acknowledgments}

We would like to thank Miranda Cheng, Paul de Lange, John Duncan, Shamit Kachru, Jared Kaplan, Ami Katz, Christoph Keller, Greg Moore, Massimo Porrati, James Sully, Erik Verlinde, Roberto Volpato, Daniel Whalen, and Xi Yin for valuable discussions, and Ami Katz for comments on a draft.  NB is supported by a Stanford Graduate Fellowship and an NSF Graduate Fellowship. ED is supported by the NSF under grant PHY-0756174 and the Department of Energy under grant DE-AC02-76SF00515.  ALF is supported by the US  Department  of  Energy Office  of  Science under Award Number DE-SC-0010025. AM is supported by the National Science and Engineering Council of Canada and by the Simons Foundation. EP is supported by the Department of Energy under Grant No. DE-FG02-91ER40671.

\begin{appendices}

\section{Special Functions and Basics of Chiral CFTs}
\subsection*{Special Functions}
\label{app:thetabohnanza}

In this appendix, we define some special functions used in the paper. We always define $q \equiv e^{2\pi i \tau}$ and $y \equiv e^{2\pi i z}$.

The Dedekind eta function is defined as
\be
\eta(\tau) = q^{\frac{1}{24}} \prod_{n=1}^{\infty} (1-q^n)
\ee
The modular discriminant is $\Delta(\tau) = \eta(\tau)^{24}$. 

The Jacobi theta functions are defined as
\begin{align}
\theta_1(\tau,z) &= -i q^{\frac18}y^\half \prod_{n=1}^{\infty}(1-q^n)(1-yq^n)(1-y^{-1}q^{n-1}) \nn \\
\theta_2(\tau,z) &= q^{\frac18}y^\half \prod_{n=1}^{\infty}(1-q^n)(1+yq^n)(1+y^{-1}q^{n-1})\nn\\
\theta_3(\tau,z) &= \prod_{n=1}^{\infty} (1-q^n)(1+yq^{n-\half})(1+y^{-1}q^{n-\half})\nn\\
\theta_4(\tau,z) &= \prod_{n=1}^{\infty} (1-q^n)(1-yq^{n-\half})(1-y^{-1}q^{n-\half}).
\end{align}
When we drop the second argument, we are implicitly setting $z$ to $0$, giving
\begin{align}
\theta_1(\tau) &= 0 \nn \\
\theta_2(\tau) &= 2q^{\frac18} \prod_{n=1}^{\infty} (1-q^n)(1+q^n)^2 \nn \\
\theta_3(\tau) &= \prod_{n=1}^{\infty} (1-q^n)(1+q^{n-\half})^2 \nn \\
\theta_4(\tau) &= \prod_{n=1}^{\infty} (1-q^n)(1-q^{n-\half})^2.
\end{align}

We also define the generators of the ring of weak Jacobi forms as
\begin{align}
&E_4(\tau) = 1 + 240\sum_{n=1}^{\infty} \sigma_3(n) q^n \nn\\
&E_6(\tau) = 1 - 504\sum_{n=1}^{\infty} \sigma_5(n) q^n \nn\\
&\phi_{0,1}(\tau,z)= 4\sum_{i=2}^4\frac{\theta_i(\tau,z)^2}{\theta_i(\tau,0)^2}\nn\\
&\phi_{-2,1}(\tau,z)=\frac{\theta_1(\tau,z)^2}{\eta(\tau)^6}
\end{align}
where
\be
\sigma_x(n) = \sum_{d|n}d^x.
\ee

Finally, we also make use of the second holomorphic Eisenstein series, $E_2(\tau)$, which we normalize as
\eq{e2def}{E_2(\tau) = 1-24\sum_{n=1}^{\infty}\sigma_1(n) q^n}
This obeys
\eq{}{q \p_q \log \D(\tau) = E_2(\tau)}
\subsection*{Basics of Chiral CFTs}\label{chiapp}

Here we provide some background on chiral CFTs. Consider a general chiral CFT with central charge $c=24k$. All chiral CFTs are theories of the identity module of some exotic $W$-algebras: every operator, being holomorphic, is either a primary current or a descendant of a primary current. 

The conventional expansion of $Z(\tau)$ in terms of characters, as in \eqr{ztau}, makes manifest the usual organization into conformal families, but obscures some crucial underlying structure. 
The theory must be conformally invariant in Euclidean signature, so that
\be
Z(\tau) = Z(\gamma\tau),~~~~~\gamma\tau \equiv {a\tau + b \over c \tau + d}
\ee
where 
\be
\gamma = \left({a~b\atop c~d}\right)\in SL(2,\Z)
\ee
describes a general modular transformation. We will denote by ${\cal F}_{SL(2,\Z)}$ the usual fundamental domain under modular transformations.  The partition function should also be finite for every value of $\tau$ with $\text{Im~}\tau>0$.  Finally, as the theory is chiral the partition function should depend holomorphically on $\tau$.
Thus the partition function can be regarded as a meromorphic function on ${\cal F}_{SL(2,\Z)}$ with a pole at $\tau=i\infty$ of order $k$.

This strongly constrains the possible chiral spectra.  One convenient way to write these constraints is to use the $j$-function $j(\tau)$, which is (up to a constant) the unique meromorphic function which maps  the fundamental domain ${\cal F}_{SL(2,\Z)}$ onto the Riemann sphere $\C^*$ with a simple pole at $i\infty$. We will choose a normalization such that 
\be
j(\tau) = q^{-1} + 744 + 196884 q + \dots
\ee
The partition function can be viewed as a meromorphic function on $\C^*$ with a $k^{\text{th}}$ order pole at $i\infty$.  Thus it is a polynomial, 
\be\label{Zj}
Z(\tau) = a_0 j(\tau)^k + a_1 j(\tau)^{k-1} +\dots a_k
\ee
for some $a_i\in\Z$. Three features of chiral CFTs are immediately apparent:
\begin{itemize}
\item
The central charge $c=24k$ is quantized, $k\in \Z$.  
\item
The dimensions $\Delta\in \Z$ are also quantized, $h\in\Z$.
\item
The partition function is uniquely determined by the $k+1$ coefficients $a_i$.
\end{itemize}
This latter point is important: it means, for example, that $Z(\tau)$ is uniquely determined by its polar part.

We note that the above results can be formulated in terms of modular functions as follows. Let us write
\be
Z(\tau) = \Delta(\tau)^{-k} {\cal Z} (\tau) 
\ee
The advantage of this definition is that ${\cal Z}(\tau)$ does not have a pole in the upper half complex $\tau$ plane. Thus, ${\cal Z}(\tau)$ is a modular form of weight $12k$, which may be expanded in terms of the Eisenstein series $E_4(\tau)$ and $E_6(\tau)$, which form a basis for the vector space of modular forms and transform with weight $4$ and $6$, respectively:
\be
{\cal Z}(\tau) = b_0 E_4^{3k} + b_1 E_4^{3k-3}E_6^2 + \dots + b_k E_6^{3k}  
\ee
We note that the number of modular forms of weight $12k$ is precisely $k+1$, and that the coefficients $b_i$ can be mapped onto the coefficients $a_i$ above.

\section{Generating Function for $\beta_{k^*}$}
\label{app:betaksGen}

In this appendix, we will derive the generating function for $\beta_{k^*}$ given in equation (\ref{eq:generatingbeta}). This derivation mirrors very closely techniques used in Section 5.2 of \cite{Gaberdiel:2008xb}.

We first define a function counting $\mc{N}=1$ descendants of the vacuum, namely 
\be
\chi^{{\cal N}=1}_{\rm vac}(\tau) = q^{-k^*\over 2}\chi(\tau)~,\quad  \text{where}\quad \chi(\tau) \equiv \prod_{n=1}^{\infty} \frac{1+q^{n+\half}}{1-q^{n+1}} = \sum_{\substack{h \geq 0 \\ h \in \half \bb{Z}}} d_h^{\mc{N}=1} q^h.
\ee
Furthermore, for each positive integer $k^*$ let us define $Z^W_{k^*}(\tau)$ to be the unique $\Gamma_{\theta}$-invariant function such that
\be
Z_{k^*}^W(\tau) = q^{-\frac{k^*}2} \chi(\tau) + \mc{O}(q^{\half}).
\ee
The $Z_{k^*}^W$ are precisely the $\mc{N}=1$ NS partition functions considered in \cite{Witten:2007kt}. $\beta_{k^*}$ is then given by
\be
\beta_{k^*} = (-1)^{k^*} Z_{k^*}^W(\tau=1).
\label{eq:sushi}
\ee

Since $Z_{k^*}^W(\tau)$ is invariant under $\Gamma_{\theta}$ and only has a pole at $\tau = i\infty$, it can be written as a polynomial in $K(\tau)$ of degree $k^*$, where $K(\tau)$ is a function parametrizing $\mc{H}/\Gamma_{\theta}$, namely
\begin{align}
K(\tau) &\equiv \frac{\Delta^2(\tau)}{\Delta(2\tau)\Delta(\tau/2)}\nn \\
&= q^{-\half} + 24 + 276q^{\half} + 2048q + 11202q^{\frac32} + \ldots.
\end{align}
We can build $Z_{k^*}^W(\tau)$ in an explicitly $\Gamma_\theta$-invariant manner from its polar pieces. In particular, for $h \in \half \bb{Z}$, let $\wp_h$ be the unique polynomial (of degree $2h$) such that
\be
\wp_h(K) = q^{-h} + \mc{O}(q^\half).
\ee
Then we have
\be
Z_{k^*}^W(\tau) = \sum_{\substack{h=0\\h\in\half\bb{Z}}}^{\frac{k^*}2} d_{h}^{\mc{N}=1} \wp_{\frac{k^*}2-h}(K).
\ee
From (\ref{eq:sushi}) we then get
\begin{align}
(-1)^{k^*} \beta_{k^*} &= \sum_{\substack{h=0\\h\in\half\bb{Z}}}^{\frac{k^*}2} d_{h}^{\mc{N}=1} \wp_{\frac{k^*}2-h}(K(\tau=1)) \nn \\
&= \sum_{\substack{h=0\\h\in\half\bb{Z}}}^{\frac{k^*}2} d_{h}^{\mc{N}=1} \wp_{\frac{k^*}2-h}(0)
\label{eq:salmon}
\end{align}
where we used the fact that $K(\tau=1)=0$. Then from (\ref{eq:salmon}) we see
\begin{align}
\sum_{k^*=1}^{\infty} (-1)^{k^*} \beta_{k^*} q^{\frac{k^*}2} &= \(\sum^{k^*\over 2}_{\substack{h=0\\h\in\half\bb{Z}}}d_h^{\mc{N}=1}q^h\)\(\sum_{n\geq0}q^{\frac{n}2} \wp_{\frac{n}{2}}(0)\) \nn \\
&= \chi(\tau)\(\sum_{n\geq0}q^{\frac{n}2} \wp_{\frac{n}{2}}(0)\).
\end{align}
In Section 5.2 of \cite{Gaberdiel:2008xb} it was shown that
\be
\sum_{n\geq0}q^{\frac{n}2} \wp_{\frac{n}{2}}(0)=\th_4(\tau)^4-\th_2(\tau)^4
\ee
which gives
\begin{align}
\sum_{k^*=1}^{\infty} (-1)^{k^*} \beta_{k^*} q^{\frac{k^*}2} &= \chi(\tau) \(\th_4(\tau)^4-\th_2(\tau)^4\) \nn \\
&= \sqrt{\frac{\theta_3(\tau)}{\eta(\tau)^3}}\(\theta_4(\tau)^4-\theta_2(\tau)^4\)q^{\frac1{16}}\(1-\sqrt{q}\).
\end{align}

\section{Perturbative Contributions to the Entropy}
\label{app:N1Sub}

There will generally be additional corrections to $S_{\rm BH}(h)$ even in perturbation theory, which will be suppressed by powers of $1/k$. Here we will compute these corrections by saddle point, and at $h=0$ obtain an expression resumming these perturbative corrections to all orders.  We first require an analytic expression for the number of descendants $d_h$ that can be used to do the saddle point integral.  For the non-supersymmetric case, (\ref{eq:DescFromPart}) can be evaluated using a famous result from Rademacher:\footnote{See e.g. \cite{rademacher1937convergent} for the original derivation, or \cite{andrews1998theory} pg 69 for a more recent treatment.}
\be
p(n) = 2\pi  \left( \frac{\frac{1}{24}}{n-\frac{1}{24}} \right)^{\frac{3}{4}}  \sum_{k=1}^\infty \frac{1}{k} A_k(n-1) \,I_{\frac{3}{2}} \left( \frac{ \pi}{k} \sqrt{ \frac{2}{3} \left( n - \frac{1}{24} \right)}  \right) ,
\label{eq:RademacherPartitions}
\ee
where 
\eq{}{A_k(n-1) = \sum^k_{m=1 \atop (m,k) =1} e^{-\pi i \left(s(m,k) - \frac{1}{k} 2n m\right)}}
with
\eq{}{s(m,k) = \sum_{n =1}^{k-1} \frac{n}{k} \left( \frac{m n}{k} - \lfloor \frac{m n}{k} \rfloor - \frac{1}{2} \right)}
We explain where this expression comes from in Appendix \ref{app:MultSys}. At large $m$, the $k>1$ terms are exponentially suppressed. Isolating the $k=1$ term and using the large argument asymptotics for the Bessel function, 
\eq{}{I_{3\over 2}(x\gg 1) \sim {e^x\over \sqrt{2\pi x}}\left(1-{1\over x}\right)~,}
gives the following approximation for the number of Virasoro vacuum descendants at level $m\gg 1$:
\be
d_{m} =\frac{2 \sqrt{3} e^{\frac{1}{6} \pi  \sqrt{24 m-1}} \left(\pi  \sqrt{24
   m-1}-6\right)}{\pi  (24 m-1)^{3/2}}-\frac{2 \sqrt{3} e^{\frac{1}{6} \pi  \sqrt{24
   m-25}} \left(\pi  \sqrt{24 m-25}-6\right)}{\pi  (24 m-25)^{3/2}},
   \label{eq:N0descapp}
      \ee
Appendix \ref{app:MultSys} applies methods of (\ref{eq:RademacherPartitions}) to obtain similar results that we will need to treat supersymmetric cases later on in this appendix.  
      
      It turns out that there is a very simple method to perform the integral approximation to (\ref{eq:OneLoop}) to all orders in $1/k$ immediately.  However, to make the logic of the argument clear, in the next subsection \ref{app:saddlehard}, we will first perform a brute force evaluation, order by order in $1/k$, which is more difficult computationally but simpler conceptually.  Then, in subsection \ref{app:saddleeasy}, we will present the simpler computation. We also extend the derivation to the $\mc{N}=1$ case.

\subsection{Direct Saddle Point Evaluation}
\label{app:saddlehard}

 Here we will compute the first few subleading corrections to the entropy at $h=0$ for the non-supersymmetric case. First, from the asymptotic formula for the number of vacuum descendants, we have
\be
\log \left( d_{x} \right) \approx \pi \sqrt{\frac{2x}{3}} - \frac{3}{2} \log x +\log\left[  \frac{\pi}{12 \sqrt{2}} \right] .
\ee
Similarly, for the Bessel function we have
\be
\log(I_1(x) )  \approx x - \frac{1}{2} \log x- \frac{1}{2} \log 2 \pi.
\ee
The integral we want to do by saddle point now includes these additional terms in its exponential, as well as a $\log 2 \pi$ from the prefactor.  Denoting this exponential by $S$, we have
\be
S \approx \pi  \left(\sqrt{\frac{2(k- |h'|)}{3}}+4 \sqrt{h |h'|}\right)+\frac{1}{4} \log
   \left(\frac{|h'|}{h^3 \left(k-|h'|\right)^6 2^{12}3^4}\right).
   \ee
   We can find the saddle point in an expansion around $k^*, |h'|$ large.  The position of the saddle point is shifted by the higher order corrections:
      \be
   h' \approx -\frac{24 h k}{1 + 24 h} - \frac{(1+144 h) \sqrt{3 k}}{\sqrt{2} (1+24 h)^{3/2} \pi}.
   \label{eq:N0saddlepoint}
   \ee
Expanding around this saddle point and including the $\log$ terms that arise from the integral, we finally arrive at
\be
S_{\rm BH}(h) \approx  \pi \sqrt{k  \left(\frac{2}{3}+ 16 h\right)} - \frac{1}{2} \log k + \log\left( \sqrt{2} \pi \right)+ \dots,
\ee
producing the result given in the second line of equation (\ref{eq:N0log}). One can continue to perform the saddle point integration by brute force to higher and higher orders, and we have explicitly checked up to ${\cal O}(k^{-2})$ that this reproduces the perturbative expansion of the first line of (\ref{eq:N0log}).

\subsection{Indirect Saddle Point Evaluation}
\label{app:saddleeasy}

Now, we will argue that the saddle point integral can be performed in closed form to all orders in the large $k$ expansion of the exponent.  Let us write the saddle point integral we are trying to do as
\be
\begin{aligned}
\int_{-\infty}^\infty M_h(h') dh' & \equiv L_h, \\
M_h(h') & = 2\pi C_{h'} \sqrt{\frac{-h'}{h}}  I_1\(4\pi \sqrt{-h' h}\) . 
\end{aligned}
\ee
As written, the integration makes sense only as a formal expression that should be interpreted as an instruction to perform the integral by saddle point.  The reason this is necessary is that $M_h(h')$ does not vanish at the boundaries of integration $h'=\pm \infty$ (in fact as $h' \rightarrow -\infty$ it grows without bound), and furthermore these boundaries are not connected to the saddle point by a path of stationary phase.    To make the integral converge, we have to regulate it:
\be
\int_{\Lambda_-}^{\Lambda_+} M_h(h') dh' \equiv L_h + S_h,
\label{eq:intreg}
\ee
where as a result of the regularization, there is now a small (i.e. exponentially subleading at large $k$) correction term $S_h$. The lower limit of integration $\Lambda_-$ can be taken to be approximately $\Lambda_- \sim -k$; the exact value does not matter because throughout our analysis, ${\cal O}(1)$ changes in $\Lambda_-$ will change the result of the integral by an exponentially subleading amount.  Naively, one might think the same is true of $\Lambda_+$; however when we take the limit $h\rightarrow 0$, small deviations in $\Lambda_+$ {\it will} matter.  To make the saddle point integration a valid approximation, we want to take $\Lambda_+$ so that the integration contour lies on a path of stationary phase. Since the path of stationary phase comes to an endpoint at $h'=0$, this restricts $\Lambda_+ \leq 0$.  Furthermore, we need the the contour to include the saddle point, so we must take $\Lambda_+ \geq -\frac{24 h k}{1+24 h} $ (from equation (\ref{eq:N0saddlepoint})).  This implies that we must take $\Lambda_+=0$ if we want $S_h$ in (\ref{eq:intreg}) to remain exponentially small even as $h\rightarrow 0$.  As a result, we can write
\be
L_0  \equiv \lim_{h\rightarrow 0} L_h = \lim_{h\rightarrow 0} \left[ \int_{\Lambda_-}^{0} M_h(h') dh' \right] + S',
\ee
where $S'$ is exponentially small and so can be discarded.  Taking the limit inside the integral, 
\be 
\begin{aligned} 
L_0 & \cong \int_{\Lambda_-}^{0} dh' M_0(h')  = \int_{\Lambda_-}^{0} dh' (-4\pi^2) h' C_{h'}\\
&= \int_{\Lambda_-}^0 dh' (-4\pi^2) h' d_{h'+k} \\
&\cong 2 \sqrt{3} \left[ e^{ \frac{\pi}{6} \sqrt{24 k-1} } - e^{\frac{\pi}{6} \sqrt{24 k-25}} \right],
\end{aligned}
\label{eq:getthreshhold}
\ee
where we have used the expression (\ref{eq:N0descapp}) for the number of descendants, and discarded the exponentially subleading $\Lambda_-$-dependent terms $\sim e^{ \frac{\pi}{6} \sqrt{24 (k+\Lambda_-) -1}}$. This is the expression given in the first line of (\ref{eq:N0log}).\footnote{Let us comment in some more detail on the structure of the saddle point integration, and in particular see why we can use just the leading order saddle point. Rescaling $h'= k x$ and taking $x_* k $ to be the leading order saddle point value, i.e. $x_* \equiv \frac{24 h}{1+24 h}$, the integrand $M_h$ has an expansion of the form
\be
M_h(k x) = \exp\left( \sqrt{k} b_0+ b_1 (x-x^*) + \sqrt{k} \Big[ -b_2 (x-x_*)^2 +b_3 (x-x_*)^3 + b_4 (x-x_*)^4 + \dots\Big]\right).
\ee
The $b_i$ coefficients are $k$- and $h$-dependent, but the leading $k$-dependence is factored out so that they behave like ${\cal O}(k^0)$ at $k\rightarrow \infty$. Consequently, all terms other than the constant term $b_0$ and the quadratic term are small and can be treated in a perturbative expansion, as one can see explicitly in the coordinates $u = k^{1/4}(x-x_*)$. Thus, as long as the lower bound of integration on $x$ is less than $x_*$, the saddle point integral is exponentially insensitive to its exact value. By taking the lower bound to be 0, it remains less than $x_*$ for all $h>0$, and it remains on the path of stationary phase.}

We can do the same procedure for the $\mc{N}=1$ case. We will show in Appendix \ref{app:MultSys} that the growth of $\mc{N}=1$ descendants of the vacuum at level $m$ goes as
\be
d_{m}^{\mathcal{N}=1}\approx \frac{\sqrt{2}}{\pi} \left( \frac{e^{\frac{\pi}{4} \sqrt{16m-1}} (\pi \sqrt{16m -1}-4)}{ (16m-1)^{3/2}} -\frac{e^{\frac{\pi}{4} \sqrt{16m-9}} (\pi \sqrt{16m -9}-4)}{(16m-9)^{3/2}} \right).
\ee
We want to integrate this against a Bessel function as in (\ref{eq:SugraOneLoop}), and take the $h\rightarrow0$ limit inside the integral. Using the fact that at small argument, $I_1(x) \sim \frac{x}2$, this gives
\begin{align}
N_{\text{BH}}^{\mc{N}=1}(0) &\cong 2(2\pi^2) \int_{\Lambda_-}^{0} dh' \(-h' d_{\frac{k^*}2+h'}^{\mc{N}=1}\) \nn \\
&\cong \sqrt{2}\left( e^{\pi  \sqrt{\frac{k^{*}}{2}-\frac{1}{16}}}- e^{\pi  \sqrt{\frac{k^{*}}{2}-\frac{9}{16}}}\right) \nn \\
S_{\text{BH}}^{\mc{N}=1}(0) &= \pi\sqrt{\frac{k^*}2} - \half\log k^* + \log\(\frac{\pi}2\) + \mc{O}\(\frac1{\sqrt{k^*}}\).
\end{align}
(Note that we get an extra factor of 2 when converting the sum in (\ref{eq:SugraOneLoop}) to an integral since the sum goes over half-integers.)

\section{Rademacher Sums and Ramanujan's Sum}
\label{app:MultSys}
Throughout the previous sections we used perturbative expansions for the coefficients of various functions expanded in $q=e^{2\pi i \tau}$. In this appendix we describe a useful technique, due to Rademacher, to arrive at these expressions.  The final expressions we use are summarized at the end of this appendix.

In 1937, Rademacher provided a nice formula for how to complete the polar part of a modular form of negative or zero weight \cite{Rademacher:1968796}. Given a modular form of weight $w<0$ with multiplier system, $\epsilon$.
\es{modf}{
f\left(\frac{a\tau+b}{c\tau+d}\right)&=\epsilon(a,b,c,d)(c\tau+d)^{w}f(\tau)\,,
}
Rademacher's formula gives the contribution to the $q^{m+\alpha}$ term from a singularity of order $\frac{1}{q^{n-\alpha}}$ as
\es{RadSum}{
a_{m,n}&=2\pi \sum_{c=1}^{\infty}\frac{A_{c,n}^{\alpha}(m)}{c}\left(\frac{n-\alpha}{m+\alpha}\right)^{\frac{1-w}{2}}I_{1-w}\(\frac{4\pi}{c}\sqrt{(n-\alpha)(m+\alpha)}\)\,.
}
Here, $A^\alpha_{c,n}(m)$ is a minor generalization of the Kloosterman sum to accommodate the multiplier system,
\es{Adef}{
A_{c,n}^{\alpha}(m)&=\sum_{\substack{d=1\\ (c,d)\,=1}}^{c}(-i)^{w}\frac{e^{-2\pi i \left(\frac{a}{c}(n-\alpha)-\frac{d}{c}(m+\alpha)\right)}}{\epsilon(a,b,c,d)}\,.
}
A classic application of this formula is to the partition of integers. Recall,
\es{partgen}{
\sum_{n=0}^{\infty}p(n)q^{n}&=\frac{q^{1/24}}{\eta(\tau)}\,.
}
Here $\eta$ is the Dedekind eta function, which transforms as,
\es{eta}{
\eta\left(\frac{a\tau+b}{c\tau+d}\right)&=(c\tau+d)^{1/2}\epsilon(a,b,c,d)\eta(\tau)\,,
}
with
\es{epeta}{
\epsilon(a,b,c,d)&=\left\{\begin{array}{cl}
e^{\frac{\pi i b}{12}} & c=0\\
\sqrt{-i} e^{\pi i  \left(\frac{a+d}{12 c}-s(d,c)\right)} & c > 0
\end{array}\right.\,,
}
and $s(d,c)$ is the Dedekind sum,
\es{dedsum}{
s(d,c)&=\sum_{n=1}^{c-1}\frac{n}{c}\left( \frac{d n}{c}-\left\lfloor{\frac{d n}{c}}\right\rfloor-\frac{1}{2}\right)\,.
}
So, $\eta(\tau)^{-1}$ is a modular form of weight $-\half$ and multiplier system $\epsilon^{-1}$, which behaves as $\lim_{\tau\rightarrow i\infty}\eta(\tau)^{-1}=q^{-1/24}+\textrm{nonpolar}$. We can thus use (\ref{RadSum}) to write,
\es{partitionsrad}{
p(\ell)&=2\pi \sum_{c=1}^{\infty}\frac{A_{c,1}^{23/24}(\ell-1)}{c}\frac{I_{3/2}\left(\frac{\pi}{6c}\sqrt{24\ell-1}\right)}{(24 \ell-1)^{3/4}}\\
&=\frac{1}{\pi\sqrt{2}}\sum_{c=1}^{\infty}A_{c,1}^{23/24}(\ell-1)\frac{d}{d\ell}\left(\frac{\sinh\left(\frac{\pi}{c}\sqrt{\frac{2}{3}(\ell-\frac{1}{24})}\right)}{\sqrt{\frac{\ell-\frac{1}{24}}{c}}}\right)\,.
}
This is Rademacher's formula for the partition of integers quoted above, (\ref{eq:RademacherPartitions}), where we simplified notation and wrote $A_{c,1}^{23/24}(n-1)$ as $A_{c}(n-1)$.

As mentioned above, we can also use the Rademacher sum to express the partition function of a chiral CFT in terms of its poles. Because the partition function is completely invariant under modular transformation, the multiplier system, weight, and shift $\alpha$ all vanish in $A^\alpha_{c,n}(m)$, so (\ref{Adef}) simplifies to\footnote{We drop the overall minus sign in the exponential of (\ref{eq:beefstroganoff}) to match conventions in the literature; when $\epsilon$, $w$, and $\alpha$ all vanish it can be shown that this minus sign does not matter after doing the sum over $d$.}
\be
A_{c,n}(m) = \sum_{\substack{d=1\\ (c,d)\,=1}}^c e^{2\pi i\(\frac acn-\frac dcm\)}.
\label{eq:beefstroganoff}
\ee
Given a pole of order $n$ we get a contribution to the $q^{m}$ term of,
\es{radpart}{
a_{m,n}&=2\pi \sum_{c=1}^{\infty}\frac{A_{c,n}(m)}{c}\sqrt{\frac{n}{m}}I_{1}\(\frac{4\pi}{c}\sqrt{n m}\)\,.
}
This can be naively continued to $m=0$ giving,
\es{m0dog}{
a_{0,n}&=4\pi^{2}n \sum_{c=1}^{\infty}\frac{A_{c,n}(0)}{c^{2}}\\
&=24\sigma_{1}(n)\,.
}
Here,
\es{radssumx}{
A_{c,n}(0)&=\sum_{\substack{a=1\\ (a,c)\,=1}}^{c}e^{2\pi i\frac{a}{c}n}\,.
}
This is known as Ramanujan's sum.
\subsection{$\Gamma_{\theta}$ Rademacher Sum}
We also would like to use Rademacher sums to get expressions for the partition functions, descendants, and $\beta_{k^{*}}$ in the supersymmetric cases. The partition function discussion is almost identical except that we now have four sectors corresponding to the four spin structures on the torus. Three of these transform into each other under $SL(2,\mathbb{Z})$, so we can either perform the appropriate Rademacher sum for vector valued modular forms, or we can consider subgroups of $SL(2,\mathbb{Z})$ which preserve the appropriate spin structure. For brevity, we will discuss the latter method, focusing on the NS sector partition function which is invariant under $\Gamma_{\theta}$,
\es{Gthdef}{
\Gamma_{\theta}&=\left\{\gamma\,=\,\mat{cc}{a&b\\c&d} \, \in \, SL(2,\mathbb{Z}) \ \ \ : a+b \textrm{ is odd and } c+d \textrm{ is odd}\right\}\,. 
}
It is easy to check that the different methods give identical results. 
The Rademacher sum for a $\Gamma_{\theta}$ modular form has coefficients given by,
\es{Gthrad}{
a_{m,n}^{(\mathcal{N}=1)}&=\pi \sum_{c=1}^{\infty}\frac{A_{c,n}^{(\mathcal{N}=1)\alpha}(m)}{c}\left(\frac{n-\alpha}{m+\alpha}\right)^{\frac{1-w}{2}}I_{1-w}\(\frac{4\pi}{c}\sqrt{(n-\alpha)(m+\alpha)}\)\,,
}
with,
\es{AN1def}{
A_{c,n}^{(\mathcal{N}=1)\alpha}(m)&=\sum_{\substack{d=1\\ (c,d)\,=1\\ c+d\,\in\,2\mathbb{Z}+1}}^{2c}(-i)^{w}\frac{e^{-2\pi i \left(\frac{a}{c}(n-\alpha)-\frac{d}{c}(m+\alpha)\right)}}{\epsilon(a,b,c,d)}\,.
}
In (\ref{AN1def}), the Kloosterman sum runs over elements of $\Gamma_{\theta}$ rather than $SL(2,\mathbb{Z})$. 

For the partition function, a pole of order $n$ contributes 
\eq{Gthpart}{
a^{(\mathcal{N}=1)}_{m,n}=\pi \sum_{c=1}^{\infty}\frac{A_{c,n}^{(\mathcal{N}=1)}(m)}{c}\sqrt{\frac{n}{m}}I_{1}\(\frac{4\pi}{c}\sqrt{n m}\)\,.
}
to the $q^{m}$ term. Again this can be continued to $m=0$ giving a slight modification of Ramanujan's sum,
\es{m0cat}{
a^{(\mathcal{N}=1)}_{0,n}&=2\pi^{2}n \sum_{c=1}^{\infty}\frac{A_{c,n}^{(\mathcal{N}=1)}(0)}{c^{2}}\\
&=8f(n)\,,
}
where we have introduced
\eq{n1div}{
f(h')=|(\text{sum of even divisors of~}h')-(\text{sum of odd divisors of~}h')|.
}
Here,
\eq{radssumblah}{
A_{c,n}^{(\mathcal{N}=1)}(0)=\sum_{\substack{d=1 \\ (d,c)=1\\d+c\in2\mathbb{Z}+1}}^{2c}e^{2\pi i\frac{a}{c}n}\,.
}

The generating function for the number of vacuum descendants for the $\mathcal{N}=1$ theory is given by
\es{n1vacdes}{
\sum_{\substack{m=0\\m\in\half\bb{Z}}}^{\infty} d_{m}^{\mathcal{N}=1}q^{m}&=q^{\frac{1}{16}}(1-\sqrt{q})\sqrt{\frac{\theta_{3}(\tau)}{\eta(\tau)^3}}\,.
}
while $\beta_{k^{*}}$ has the generating function,
\es{betaksgen}{
\sum_{k^{*}=1}^{\infty}(-1)^{k^{*}}\beta_{k^{*}}q^{k^{*}/2}&=q^{\frac{1}{16}}(1-\sqrt{q})\sqrt{\frac{\theta_{3}(\tau)}{\eta(\tau)^3}}(\theta_{4}(\tau)^4-\theta_{2}(\tau)^4)\,.
}
as derived in Appendix \ref{app:betaksGen}. Ignoring the prefactor of $q^{\frac1{16}}(1-\sqrt{q})$, these both have nice modular properties. Although they do not have negative weight and the Rademacher sum (\ref{Gthrad}) does not give a convergent expansion for the coefficients, it does still give an asymptotic expansion from which we can read off the perturbative part from the $c=1$ term in the sum. Afterwards, we can easily convolve back in the $q^{\frac{1}{16}}(1-\sqrt{q})$ term. This gives
\hspace{-.1\textwidth}\es{n1descpert}{
\hspace{-.1\textwidth}d_{m}^{\mathcal{N}=1}&\approx \frac{\sqrt{2}}{\pi} \left( \frac{e^{\frac{\pi}{4} \sqrt{16m-1}} (\pi \sqrt{16m -1}-4)}{(16m-1)^{3/2}} -\frac{e^{\frac{\pi}{4} \sqrt{16m-9}} (\pi \sqrt{16m -9}-4)}{(16m-9)^{3/2}} \right) \,, \\
\hspace{-.3\textwidth}\log\left(d_{m}^{\mathcal{N}=1}\right)&\approx \pi \sqrt{m} - \frac32\log m + \log\( \frac{\pi}{32\sqrt{2}}\)+\ldots\,,
}
and,
\es{bkdescpert}{
(-1)^{k^{*}+1}\beta_{k^{*}}&\approx\sqrt{2}\left( e^{\pi  \sqrt{\frac{k^{*}}{2}-\frac{1}{16}}}- e^{\pi  \sqrt{\frac{k^{*}}{2}-\frac{9}{16}}}\right)\,,\\
\log\left((-1)^{k^{*}+1}\beta_{k^{*}}\right)&\approx \frac{\pi  \sqrt{k^{*}}}{\sqrt{2}}+\log \left(\frac{\pi }{2 \sqrt{k^{*}}}\right)\,.
}

We can attempt to repeat the same calculation using the (ungraded) $\mc{N}=2$ vacuum descendants instead (see subsection \ref{sec:ungrade}). The generating function for the $\mc{N}=2$ vacuum descendants is
\be
\sum_{\substack{m=0 \\ m \in \half\bb Z}}^{\infty} d_m^{\mc{N}=2} q^m = \frac{1-\sqrt{q}}{1+\sqrt{q}} q^{\frac18} \frac{\theta_3(\tau)}{\eta(\tau)^3}
\label{eq:thing1}
\ee
while $\beta_{k^*}^{\mc{N}=2}$ has the generating function
\be
\sum_{k^*=1}^{\infty} (-1)^{k^*} \beta_{k^*}^{\mc{N}=2} q^{k^*/2} =  \frac{1-\sqrt{q}}{1+\sqrt{q}} q^{\frac18} \frac{\theta_3(\tau)}{\eta(\tau)^3}(\theta_{4}(\tau)^4-\theta_{2}(\tau)^4).
\label{eq:thing2}
\ee
Unlike in the $\mc{N}=1$ case, however, the prefactor to a nice modular object is now $\frac{1-\sqrt{q}}{1+\sqrt{q}} q^{\frac18}$. Now in order to get the growth of either $d_m^{\mc{N}=2}$ or $\beta_{k^*}^{\mc{N}=2}$, we would need to convolve against $k^*$ terms, and we do not know how to do this sum in closed form. We can, however, numerically approximate the growth of both of these functions. In doing so, we get
\begin{align}
\log\(d_m^{\mc{N}=2}\) &\approx \sqrt{2}\pi\sqrt{m} - \frac74 \log{m} + \(\log \pi - \frac{25\log{2}}4\) - \( \frac{35+\pi^2}{8\sqrt{2}\pi} \)\frac{1}{\sqrt{m}}  \nn \\ &+ \frac{0.00512158}m + \frac{0.12632989}{m^{3/2}}-\frac{0.13532569}{m^2} + \mc{O}\(\frac{1}{m^{5/2}}\) \nn \\
\log\((-1)^{k^*+1}\beta_{k^*}^{\mc{N}=2}\) &\approx \pi\sqrt{k^*} - \frac{3}{4} \log{k^*} - \log{\(\frac{4\sqrt2}{\pi}\)} - \(\frac{3+\pi^2}{8\pi}\)\frac{1}{\sqrt{k^*}} - \frac{0.03711448}{k^*} \nn \\ &+ \frac{0.21704365}{(k^*)^{3/2}}-\frac{0.13028682}{(k^*)^2}+\mc{O}\(\frac{1}{(k^*)^{5/2}}\)
\end{align}
When we take the expression for $\log\(d_m^{\mc{N}=2}\)$ above, and use methods in Appendix \ref{app:saddlehard} to find the number of states $\frac{k^*}2$ above the vacuum, we indeed reproduce exactly $\log\((-1)^{k^*+1}\beta_{k^*}^{\mc{N}=2}\)$ above.

\subsection{Summary}

For convenience, here we gather the key expressions derived in this appendix:
\begin{tcolorbox}
\begin{equation}
\begin{aligned}
\hspace{-.1\textwidth}d_{m}^{\mathcal{N}=0}&\approx \frac{2\sqrt{3}}{\pi} \left( \frac{e^{\frac{\pi}{6} \sqrt{24m-1}} (\pi \sqrt{24m -1}-6)}{ (24m-1)^{3/2}} -\frac{e^{\frac{\pi}{6} \sqrt{24m-25}} (\pi \sqrt{24m -25}-6)}{ (24m-25)^{3/2}} \right) \,,\\
\hspace{-.1\textwidth}d_{m}^{\mathcal{N}=1}&\approx \frac{\sqrt{2}}{\pi} \left( \frac{e^{\frac{\pi}{4} \sqrt{16m-1}} (\pi \sqrt{16m -1}-4)}{ (16m-1)^{3/2}} -\frac{e^{\frac{\pi}{4} \sqrt{16m-9}} (\pi \sqrt{16m -9}-4)}{(16m-9)^{3/2}} \right) \,,\\
(-1)^{k^{*}+1}\beta_{k^{*}}&\approx\sqrt{2}\left( e^{\pi  \sqrt{\frac{k^{*}}{2}-\frac{1}{16}}}- e^{\pi  \sqrt{\frac{k^{*}}{2}-\frac{9}{16}}}\right)\,,\\
\log\(d_m^{\mc{N}=2}\) &\approx \sqrt{2}\pi\sqrt{m} - \frac74 \log{m} + \(\log \pi - \frac{25\log{2}}4\) - \( \frac{35+\pi^2}{8\sqrt{2}\pi} \)\frac{1}{\sqrt{m}}  \,, \\ 
&\text{~~~}+ \frac{0.00512158}m + \frac{0.12632989}{m^{3/2}}-\frac{0.13532569}{m^2} + \mc{O}\(\frac{1}{m^{5/2}}\) \,, \\
\log\((-1)^{k^*+1}\beta_{k^*}^{\mc{N}=2}\) &\approx \pi\sqrt{k^*} - \frac{3}{4} \log{k^*} - \log{\(\frac{4\sqrt2}{\pi}\)} - \(\frac{3+\pi^2}{8\pi}\)\frac{1}{\sqrt{k^*}} - \frac{0.03711448}{k^*} \,, \\ 
&\text{~~~}+ \frac{0.21704365}{(k^*)^{3/2}}-\frac{0.13028682}{(k^*)^2}+\mc{O}\(\frac{1}{(k^*)^{5/2}}\) \,
\end{aligned}
\end{equation}
\end{tcolorbox}

\end{appendices}

\newpage

\bibliographystyle{utphys}
\bibliography{refs}

\end{document}